\documentclass[useAMS]{mn2e} \usepackage{epsfig,lscape} \usepackage{graphicx}

\DeclareGraphicsExtensions{.eps,.eps.gz,.epsi}

\newcommand\aj{AJ} \newcommand\apj{ApJ} %
\newcommand\aap{A\&A} \newcommand\pasp{PASP} \newcommand\mnras{MNRAS}%
\newcommand\apjs{ApJS}  %

\newcommand\pasa{PASA}

\newcommand{\ha}{H$\alpha$}
\newcommand{\lya}{Ly$\alpha$}
\newcommand{\hb}{H$\beta$}

\newcommand{\kms}{km s$^{-1}$}

\newcommand{\hii}{{\sc H\,II}}

\newcommand{\foiii}{[O~{\sc iii}]} 
 \newcommand{\foii}{[O~{\sc ii}]}
\newcommand{\fnii}{[N~{\sc ii}]}
\newcommand{\fsii}{[S~{\sc ii}]}

\begin{document}

\title[Hunting for Extremely Faint PNe in SDSS]{Hunting for Extremely Faint Planetary
Nebulae in the SDSS Spectroscopic Database}

\author[Yuan \& Liu]{H. B. Yuan$^{1}$\thanks{LAMOST
Fellow}\thanks{E-mail: yuanhb4861@pku.edu.cn} \&
X. W. Liu$^{1,2}$\\ \\ $1$ Kavli Institute for Astronomy and Astrophysics,
Peking University, Yi He Yuan Road 5, Hai Dian District, Beijing 100871, China \\ 
$2$ Department of Astronomy, Peking University, Yi He Yuan Road 5, Hai Dian District, Beijing 100871, China}

\date{Received:}

\pagerange{\pageref{firstpage}--\pageref{lastpage}} \pubyear{2013}

\maketitle

\label{firstpage}

\begin{abstract}{ Using $\sim$~1,700,000 target- and sky-fiber spectra from the Sloan Digital Sky
Survey (SDSS), we have carried out a systematic search for 
Galactic planetary nebulae (PNe) via  detections of the \foiii~$\lambda\lambda$4959, 5007 lines.  
Thanks to the excellent sensitivity of the SDSS spectroscopic surveys,
this is by far the deepest search for PNe ever taken,
reaching a surface brightness of the \foiii~$\lambda$5007 line 
down to about 29.0 magnitude arcsec$^{-2}$.
The search leads to the recovery of 13 previously known PNe in the Northern and Southern Galactic Caps.
In total, 44 new planetary nebula (PN) candidates are identified, including 7  
candidates of multiple detections and 37 candidates of single detection.  
The 7 candidates of multiple detections are all extremely large (between 21\arcmin~and 154\arcmin) and faint,
located mostly in the low Galactic latitude region and with a kinematics similar to disk stars.
After checking their images in \ha~and other bands, three of them are probably \hii~regions,
one is probably associated with a new supernova remnant, another one is possibly a true PN, and 
the remaining two could be either PNe or supernova remnants.
Based on sky positions and kinematics, 
7 candidates of single detection probably belong to the halo population.
If confirmed, they will increase the number of known PNe in the Galactic halo significantly. 
All the newly identified PN candidates are very faint, with a surface brightness of the \foiii~$\lambda$5007
line between 27.0 -- 30.0 magnitude arcsec$^{-2}$,
very challenging to be discovered with previously employed techniques 
(e.g. slitless spectroscopy, narrow-band imaging), 
and thus may greatly increase the number of "missing" faint PNe.
Our results demonstrate the power of large scale fiber spectroscopy in hunting for ultra-faint PNe 
and other types of emission line nebulae. Combining the large spectral databases provided by 
the SDSS and other on-going projects  (e.g. the LAMOST Galactic surveys),
it is possible to build a statistically meaningful sample of ultra-faint, large, evolved PNe,
thus improving the census of Galactic PNe.}
\end{abstract}

\section{Introduction}

Planetary nebulae (PNe) represent the last stages of evolution for low- and
intermediate-mass stars of masses less than about 8 solar
masses.  The PN phase begins when the central contracting white dwarf reaches an effective
temperature of above 30,000 K and starts to ionize the gaseous envelope ejected during
the previous Asymptotic Giant Branch (AGB) phase.  The PN phase is short,
lasting for just tens of thousands years maximally.  PNe enrich the interstellar
medium (ISM) with dust grains, helium, carbon, nitrogen, oxygen and other products of
nucleosynthesis, and are vital probes of the stellar 
nucleosynthesis processes, abundance gradients and chemical evolution of
galaxies.  In addition, PNe play a key role in studying the physics and
time-scales of mass-loss and stellar evolution for low- and intermediate-mass stars
(Iben 1995). PNe may also be progenitors of Type Ia supernovae that return
large amounts of iron to the ISM.

New PNe have been continually discovered via various surveys (Parker et al. 2012). 
The Strasbourg-ESO Catalog of Galactic Planetary Nebulae (SECGPN) 
compiled by Acker et al. (1994, 1996) contains $\sim$~1,500 true, probable or possible PNe.
Kohoutek (2001) lists $\sim$~1,500 objects classified as Galactic PNe known up to the end of 1999,
as well as a number of possible pre-PNe and post-PNe. Using the the
Anglo-Australian Observatory UK Schmidt Telescope (AAO/UKST) 
SuperCOSMOS H$\alpha$ Survey (SHS; Parker et al. 2005), Parker et al.
(2006) and Miszalski et al. (2008) present $\sim$~1,200 newly discovered true, possible or likely
Galactic PNe in the Macquarie/AAO/Strasbourg H-Alpha Planetary Nebula
Catalogue (MASH) and its supplement (MASH-II).  From the Isaac Newton Telescope (INT) Photometric
H$\alpha$ Survey (IPHAS; Drew et al. 2005), Viironen et al. (2009a, b) discover
several PN candidates.  Jacoby et al. (2010) describe a technique 
to search for additional unknown PNe via visually inspecting the existing data collections of the digital
sky surveys (DSS) such as the POSS-I and POSS-II surveys, and find tens of new PNe.  
Up to now, there are in total less than 2,850 true or probable PNe known in the Milky Way (Miszalski et al. 2012).
However, the number is still a small fraction of the total predicted by any PN
population model.  For example, the stellar population synthesis models of
Moe \& De Marco (2006) predict a total number of 46,000 $\pm$
13,000 PNe with radii $\le$ 0.9 pc in the Milky Way if  all low- and intermediate-mass 
stars of  1 –- 8 solar mass will go through a PN phase. The number drops to 
$\sim$~6,600 if all PNe have to form via the channel of close binaries 
(through the common envelope phase) (De Marco \& Moe 2005).
Observationally, based on a solar neighborhood ($\le$ 2 kpc) sample of about 200 PNe, 
Frew (2008) estimates a total Galactic 
population of 24,000 $\pm$ 4,000 PNe of radii $\le$ 1.5 pc, or 13,000 $\pm$ 
2,000 PNe of radii $\le$ 0.9 pc.

A variety of multi-waveband observations ranging from the radio
up to the X-ray have been utilized to detect PNe.
Essentially, all the efforts have been based on
the technique of wide field, interference filter or objective prism slitless spectroscopic imaging
surveys, primarily searching for emission in the light of the \foiii~$\lambda$5007 line and/or \ha.
However, it is a very challenging task to to find all PNe in the Galaxy directly. 
Firstly, since most PNe are distributed in the Galactic plane where dust extinction is 
high, a large fraction of them are unobservable at optical
wavelengths where PNe are most luminous. Secondly, when only imaging data are available, 
it is difficult to differentiate PNe from other diffuse emission line nebulae such as 
\hii~regions, supernova remnants, nova shells, symbiotic nebulae
and distant emission line galaxies.

The Sloan Digital Sky Survey (SDSS; York et al. 2000) provides uniform and
contiguous imaging photometry for about one-third of the sky in the $u,g,r,i$ and $z$ bands,
as well as over two million high quality low resolution optical spectra of stars, galaxies and quasars. 
The latest SDSS Data Release 9 (DR9; Ahn et al. 2012) delivers spectra for
about 1.46 million galaxies, 0.23 million quasars and 0.67 million stars.
The SDSS observations mainly target high Galactic latitude regions of the Northern and Southern Galactic Caps. 
If by chance  a PN (or an extended emission line nebula of other types) falls in the 
sightline of some SDSS targets (including sky background targets by sky fibers), lines emitted by the nebula, 
such as the \foiii~$\lambda\lambda$4959, 5007, typically the strongest for a medium to high excitation nebula, 
will appear in the spectra of those targets. 
Then by systematically searching and measuring the \foiii~$\lambda\lambda$4959, 5007 emission 
lines of Galactic nebular origins in those millions of SDSS spectra, one may expect to discover 
and study previously unknown PNe or other types of extended emission line nebula in the Galaxy.
This spectroscopic approach has the following advantages: 
a) Compared to previously commonly used techniques, such as narrow-band imaging and slitless 
objective prism imaging spectroscopy, the sky background is much reduced in slit or fiber spectroscopy. 
Thus the latter is much more sensitive to faint nebular emission and capable of detecting 
large (evolved and/or nearby) PNe of very low surface brightness;
b) With spectral information available over a wide wavelength coverage, it is much easier to differentiate different 
types of emission line nebulae; and c) With a contiguous and uniform coverage over a huge sky area such as 
that provided by the SDSS, it is possible to construct a statistically meaningful sample to improve the census 
of Galactic PNe.

In this paper, we present results of a systematic search for PNe 
in the SDSS spectroscopic database.
The paper is organized as following. In Section 2, we introduce the data and method used
to search for PNe. The results are presented in Section 3 and discussed in 
Section 4. A brief summary then follows in Section 5.

\section{Data and method}
\subsection{Data}
We used the SDSS DR7 (Abazajian et al. 2009) spectroscopic database in the current work.
The release contains over 1.6 million low-resolution ($R \sim 1,800$) spectra 
targeting approximately 930,000 galaxies, 120,000 quasars, 460,000 stars and 97,000 blank sky positions.
Most targets are within a large contiguous area over 7,350 deg$^2$ in the Northern Galactic Cap,
with the remaining ones from a number of stripes in the Southern Galactic Cap and 
a few stripes across the Galactic plane targeted by the program of 
the Sloan Extension for Galactic Understanding and Exploration (SEGUE; Yanny et al. 2009).

Spectroscopic observations of the SDSS are usually undertaken in non-photometric conditions when
the imaging camera is not in use. At least three fifteen-minute exposures are
taken until the cumulative mean signal-to-noise ratio (SNR) per pixel exceeds 4 for a fiducial fiber
magnitude of $g$ = 20.2 and $i$ = 19.9. For faint SEGUE plates, a total exposure 
time of about 1.5 hours is required. A total number of 640 spectra are collected simultaneously 
covering 3,800 -- 9,200\,{\AA}, at a spectral resolution of $\sim$~1,800. 
The large aperture size of the SDSS telescope (2.5 meter) and 
the long exposures ($\ge$~45 minutes) make 
the SDSS spectra extremely sensitive to narrow nebular emission lines.
Note that each SDSS fiber samples a sky area of angular diameter of 3\arcsec~
and the SDSS wavelength scale is based on vacuum wavelengths.

\subsection{Method}
For each spectrum from the SDSS DR7 spectroscopic database, we have performed 
Gaussian fitting to the \foiii~$\lambda\lambda$4959, 5007 lines around their rest wavelengths.
The two lines are fitted independently. 
To reduce the degrees of freedom of fit, we have adopted a fixed line width of FWHM 2.82\,\AA~and assumed 
a flat continuum in the vicinity of both lines. 
When a positive signal is detected, the SNR of the detection 
is then computed by dividing the peak value of the fitted Gaussian by the standard deviation of 
a segment spectrum of a wavelength span of 40\,{\AA} around the rest wavelength of the line.
Note that the SNRs thus calculated are lower limits of the true values. 
In cases of very strong signals, the SNRs would be greatly under-estimated.

To minimize spurious detections, we require that: a) SNRs of the \foiii~$\lambda$5007 line should be higher than 3;
b) The difference of radial velocities deduced from the two \foiii~$\lambda\lambda$4959, 5007 lines must be smaller than 60~\kms;
c) The intensity ratio of the two lines $F_{5007}$/$F_{4959}$ must fall between 2 and 4, 
bracketing the intrinsic ratio of 2.98 (Mathis \& Liu 1999; Storey \& Zeippen 2000).
The \foiii~$\lambda$5007 line emission from Galactic PNe may be contaminated by emission lines 
(e.g. the \foii~$\lambda\lambda$3726, 3729 lines, the Balmer lines and the \foiii~$\lambda\lambda$4959, 5007 lines)
of target galaxies and quasars. Such cases are also carefully avoided by visual check 
of the SDSS images and spectra to remove nearby galaxies whose \foiii~ emission lines are regarded as Galactic emission 
and distant galaxies and quasars that have an emission line shifted to the wavelength of the \foiii~$\lambda$5007 line.
Faint background emission line galaxies at a given redshift, such as \foii~emission line galaxies at redshift about 0.35 
and \lya~emitters at redshift about 3.1, are important contaminations for PNe surveys 
beyond the Milky Way (e.g. Gerhard 2006 and references therein). To further avoid such contaminations, 
we have excluded a few PN candidates whose \foiii~$\lambda$5007 line widths are larger than 4.23\,\AA~(1.5 times the expected line width).
These measures have lessened the extragalactic contamination in our work.
The radial velocities of the PN candidates (see Fig.\,14 in Section\,4) also suggest their Galactic origins,
as extragalactic origins would result a uniform radial velocity distribution.
In addition, we find a few cases that some spectral features of a normal galaxy at a given redshift, 
such as the 4661\AA~feature at redshift around 0.074 and the 4819\AA~feature at redshift about 0.039,
can mimic the \foiii~$\lambda$5007 emission. Such cases are also excluded.

Detection of the \foiii~$\lambda\lambda$4959, 5007 lines is a good indicatation but not sufficient to identify a PN.
The PN candidates found in this work may be contaminated by reflection nebulae, 
diffuse ionized gas, \hii~regions, supernova remnants, emission line stars and nova shells (Frew \& Parker 2010).
Imaging data, particularly narrow-band images, can provide important clues to classify the PN candidates. 
Therefore, both the SDSS spectra and available imaging data are used to help classify the nature of the PN candidates in this work.

\section{Results}
\begin{figure*}
\centering
\includegraphics[width=180mm]{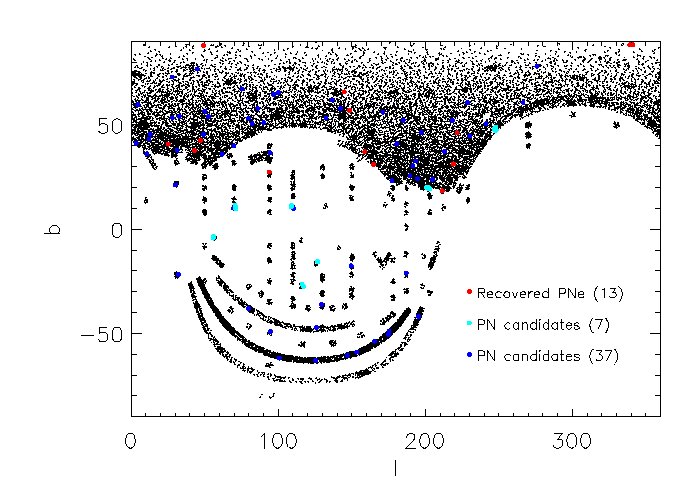}
\caption{Galactic distribution of recovered  PNe (red) within the SDSS footprint,  
newly discovered PN candidates of multiple detections (cyan) and of 
single detection (blue). Black dots show SDSS spectroscopic targets -- to avoid crowdness, only 1 per cent
randomly selected SDSS spectroscopic targets are shown.} \label{}
\end{figure*}

After applying the above criteria, a total number of 160 spectra with reliable detections 
of the Galactic \foiii~$\lambda\lambda$4959, 5007 
emission lines are selected. The spectral sequence number, SDSS spectral ID,
the type, redshift and equatorial and Galactic coordinates of the target, 
the surface brightness and radial velocity of the detected Galactic \foiii~$\lambda$5007 
line of each spectrum are listed in Tab.\,1.
Note the redshifts here indicate the SDSS targets and are given by
the SDSS pipeline. While the radial velocities are associated the PNe or PN candidates
and measured from the \foiii~$\lambda$5007 emission lines in this work.
The SDSS spectra and colour-composite thumbnails are displayed in Figs.\,A1 and A2 in the Appendix, respectively.
Based on the coordinates and radial velocities, they are divided into 58
groups, with each group representing a PN (candidate). After cross-correlating with existing PN catalogs 
including SECGPN, MASH, MASH-II, Kohoutek et al. (2001) and the IPHAS PN catalog, 
the 58 groups are further divided into 3 categories: previously known PNe or other types of emission line nebulae 
recovered in this work (Spectral sequence number SEQ 1 -- 37), 
PN candidates of multiple \foiii~$\lambda$5007 detections (SEQ 38 -- 123) and 
PN candidates of single detection (SEQ 124 -- 160), with each category having 14, 7 and 37 members, respectively.
Their Galactic distribution is shown in Fig.\,1.

As listed in Tab.\,1, the 160 SDSS spectroscopic targets include
59 galaxies (GAL), 28 blank skies (SKY), 16 blue horizontal-branch stars (BHB), 12 F-turnoff stars (FTO), 
7 quasars (QSO), 9 white dwarfs (WD), 7 X-ray sources from the ROSAT All-Sky Survey (ROS; Voges et al. 1999), 
7 G-dwarfs (GD), 3 hot stars (Hot), plus a few other types of object.
Note that the target types are assigned by the SDSS.
The fraction of spectra with detectable Galactic \foiii~line emission are 
0.007, 0.004, 0.006 and 0.016 per cent for target type GAL, QSO, GD and FTO, respectively.
The fractions increase to 0.022, 0.025 and 0.022 per cent for target type ROS, BHB and SKY,
and further to 0.087 and 0.12 for Hot and WD, respectively.
Such increases are caused by the facts that it is easier to detect the \foiii~$\lambda\lambda$4959, 5007 emission 
on spectra of blue stars or blank skies and some of the WDs and hot stars are exactly the central stars of the PNe.

\begin{table*} \begin{minipage}[]{180mm} \centering
\caption{List of SDSS targets whose spectra show detectable \foiii~$\lambda\lambda$4959, 5007 emission from Galactic PNe, \hii~regions and
PN candidates.}
\label{}
\begin{tabular}{|l|c|c|c|c|c|c|c|c|r|} \hline
SEQ & Spectr. ID$^a$ & Type$^b$  & Redshift & RA  &Dec  & l & b & S$_{5007}$$^c$ & Vr  \\ 
       &                & (Initial)         &          &(J2000.0) &(J2000.0)& &   &                &  (km s$^{-1}$)    \\ \hline       
\hline \noalign{\vskip2pt} \multicolumn{9}{c}{Recovered PNe and \hii~regions} \\
\noalign{\vskip2pt} \multicolumn{9}{l}{PN G025.3+40.8, IC\,4593, 13\arcsec, 22.0~\kms } \\
1$^d$ & 54569-2527-241 &  GAL &  0.07275 &      242.91496 &       12.04709 &       25.29303 &       40.84380 & 28.13 &   24.4 \\
\hline \noalign{\vskip2pt} \multicolumn{9}{l}{PN G043.1+37.7, NGC\,6210, 16.2\arcsec, $-$36.2~\kms} \\
2$^d$ & 53135-1414-173 &  GAL &  0.04877 &      251.23384 &       23.77270 &       43.11331 &       37.65422 & 28.19 &  $-$57.0 \\
3$^d$ & 53135-1414-174 &   MD & $-$0.00038 &      251.16985 &       23.70874 &       43.01453 &       37.69176 & 28.36 &  $-$37.1 \\
4$^d$ & 53135-1414-176 &  GAL &  0.20437 &      251.09543 &       23.76732 &       43.06049 &       37.77392 & 26.67 &  $-$40.5 \\
5$^d$ & 53135-1414-194 &  GAL &  0.43010 &      251.05946 &       23.77403 &       43.05658 &       37.80737 & 27.40 &  $-$47.6 \\
6$^d$ & 53135-1414-198 &  FTO &  2.19684 &      251.07339 &       23.78899 &       43.07936 &       37.79949 & 27.13 &  $-$39.2 \\
\hline \noalign{\vskip2pt} \multicolumn{9}{l}{PN G047.0+42.4, A\,39, 174.0\arcsec} \\
7 & 52822-1408-528 &  GAL &  0.13308 &      246.88931 &       27.88820 &       47.02350 &       42.47912 & 23.16 &    3.0 \\
8$^d$ & 52822-1408-536 &  QSO & $-$0.00038 &      246.86891 &       27.92838 &       47.07146 &       42.50535 & 27.62 &   13.5 \\
9 & 53149-1421-253 &  GAL &  0.13302 &      246.87674 &       27.91645 &       47.05768 &       42.49604 & 24.14 &    5.3 \\
\hline \noalign{\vskip2pt} \multicolumn{9}{l}{PN G049.3+88.1, H\,4-1, 2.7\arcsec, $-$141.0~\kms } \\
10$^e$ & 54156-2241-108 &  QSO &  $-$0.00062 &      194.86575 &       27.63631 &       49.30931 &       88.14751 & 18.14$^f$ &   -184.8$^f$ \\
\hline \noalign{\vskip2pt} \multicolumn{9}{l}{PN G094.0+27.4, K\,1-16, 114.0\arcsec } \\
11 & 54630-2552-347 &   WD &  0.18159 &      275.46713 &       64.36484 &       94.02541 &       27.42847 & 24.65 &  $-$58.6 \\
12 & 54632-2552-347 &   WD & $-$0.00007 &      275.46713 &       64.36484 &       94.02541 &       27.42847 & 24.65 &  $-$57.2 \\
\hline \noalign{\vskip2pt} \multicolumn{9}{l}{\hii~region, PHL\,932, 275.0\arcsec} \\
13$^g$ & 51821-0421-369 &  GAL &  $-$0.00003 &       14.97583 &       15.73003 &      125.92473 &      $-$47.09252 & 28.33 &  $-$10.5 \\
\hline \noalign{\vskip2pt} \multicolumn{9}{l}{PN G144.8+65.8, BE\,UMa} \\
14 & 52412-0968-054 &  GAL &  0.07743 &      179.43967 &       48.92075 &      144.82858 &       65.86246 & 27.08 &  $-$83.6 \\
15 & 52412-0968-060 &  QSO &  4.24529 &      179.49153 &       48.94882 &      144.72203 &       65.85540 & 27.94 & $-$127.0 \\
\hline \noalign{\vskip2pt} \multicolumn{9}{l}{PN G148.4+57.0, NGC\,3587, 170.0\arcsec, 6.0~\kms } \\
16$^d$ & 52649-1012-449 &  QSO &  1.18744 &      168.76855 &       55.02877 &      148.42288 &       57.06946 & 26.15 &   $-$9.5 \\
17 & 52649-1012-405 &  GAL &  0.00007 &      168.68710 &       55.03819 &      148.47510 &       57.03123 & 21.26 &    1.1 \\
18$^d$ & 52649-1012-406 &  GAL &  $-$0.00002 &      168.62302 &       55.01423 &      148.55469 &       57.02437 & 25.80 &   $-$6.6 \\
19 & 52707-1013-340 &  GAL &  $-$0.00001 &      168.69135 &       55.00916 &      148.50752 &       57.05441 & 20.46 &   $-$0.7 \\
\hline \noalign{\vskip2pt} \multicolumn{9}{l}{PN G158.8+37.1, A\,28, 270.0\arcsec, $-$2.0~\kms} \\
20 & 54425-1784-532 &  HOT &  $-$0.00006 &      130.39819 &       58.23010 &      158.80217 &       37.17947 & 26.91 &   $-$8.2 \\
\hline \noalign{\vskip2pt} \multicolumn{9}{l}{PN G164.8+31.1, JnEr\,1, 380.0\arcsec, $-$84~\kms} \\
21 & 53383-1870-464 &  GAL &  $-$0.00028 &      119.42754 &       53.44424 &      164.77774 &       31.16045 & 24.13 &  $-$53.9 \\
22 & 53383-1870-465 &  GAL &  $-$0.00027 &      119.42459 &       53.40979 &      164.81770 &       31.15596 & 24.91 &  $-$71.5 \\
23 & 53383-1870-466 &  GAL &  $-$0.00022 &      119.40196 &       53.43252 &      164.78998 &       31.14433 & 24.44 &  $-$66.0 \\
\hline \noalign{\vskip2pt} \multicolumn{9}{l}{PN G211.4+18.4, HDW\,7, 94.0\arcsec} \\
24 & 54505-2945-483 &   WD &  0.00058 &      118.79710 &        9.55256 &      211.46860 &       18.46137 & 25.19 &   60.1 \\
\hline \noalign{\vskip2pt} \multicolumn{9}{l}{PN G219.1+31.2, A\,31, 970.0\arcsec, 41.0~\kms} \\
25 & 53086-1760-021 &  GAL &  0.06295 &      133.63991 &        8.87866 &      219.19708 &       31.35715 & 26.37 &   25.9 \\
26 & 53086-1760-023 &  GAL &  0.06407 &      133.48776 &        8.79611 &      219.20520 &       31.18578 & 26.80 &   20.3 \\
27 & 53086-1760-026 &  GAL &  0.06426 &      133.62056 &        8.74568 &      219.32645 &       31.28074 & 27.62 &    2.3 \\
28 & 53086-1760-029 &  GAL &  0.07250 &      133.56465 &        8.85769 &      219.18030 &       31.28123 & 24.71 &   38.3 \\
29 & 53086-1760-031 &  SKY & ******* &      133.57205 &        8.92268 &      219.11600 &       31.31669 & 24.38 &   35.6 \\
30 & 53086-1760-034 &  GAL &  0.06387 &      133.56229 &        8.98637 &      219.04419 &       31.33636 & 26.49 &   20.6 \\
31 & 53086-1760-038 &  QSO &  1.64963 &      133.53564 &        8.90037 &      219.12065 &       31.27456 & 24.26 &   35.8 \\
32 & 53086-1760-039 &  GAL &  0.17061 &      133.48769 &        8.90032 &      219.09605 &       31.23210 & 25.35 &   35.1 \\
\hline
\end{tabular}
\begin{description}
\item[$^a$]  In format mjd$-$plateid$-$fiberid. 
\item[$^b$]  GAL: galaxy; MD: M-dwarf; FTO: F-turnoff star; QSO: quasar; WD: white dwarf; HOT: hot star; 
ROS: ROSAT source; GD: G-dwarf; BHB: blue horizontal branch star; KG: K-giant; SER: hand-selected target. 
\item[$^c$]  Using the definition by Jacoby (1989): $m_{5007}$=$-2.5$log($F_{5007}$)$-$13.74. 
$S_{5007}$=$m_{5007} + 2.5$log($\pi \times r^2$), $r$=1.5\arcsec~is the SDSS fiber radius.
\item[$^d$]  Indicating haloes of PNe.
\item[$^e$]  Indicating halo PNe and PN candidates.
\item[$^f$]  Based on \foiii~$\lambda$4959 line. The \foiii~$\lambda$5007 line of this target is partly clipped as cosmic rays.  
\item[$^g$]  PHL\,932 is categorised in the SECGPN catalog but in fact an \hii~region (Frew et al. 2010).
\end{description} \end{minipage} \end{table*}

\setcounter{table}{0}
\begin{table*} \begin{minipage}[]{180mm} \centering
\caption{\it -- continued}
\label{}
\begin{tabular}{|l|c|c|c|c|c|c|c|c|r|} \hline
SEQ & Spectr. ID$^a$ & Type$^b$  & Redshift & RA  &Dec  & l & b & S$_{5007}$$^c$ & Vr  \\ 
       &                & (Initial)         &          &(J2000.0) &(J2000.0)& &   &                &  (km s$^{-1}$)    \\ \hline 
\hline \noalign{\vskip2pt} \multicolumn{9}{l}{PN G221.5+46.3, EGB\,6, 720.0\arcsec} \\
33 & 54153-2584-288 &  GAL &  0.10275 &      148.24554 &       13.75261 &      221.57625 &       46.36752 & 27.92 &  $-$30.2 \\
34 & 54153-2584-297 &  GAL &  0.24429 &      148.22514 &       13.75945 &      221.55511 &       46.35248 & 27.42 &    0.2 \\
\hline \noalign{\vskip2pt} \multicolumn{9}{l}{PN G339.9+88.4, LoTr\,5, 525.0\arcsec} \\
35 & 54505-2661-581 &  SKY & ******* &      193.89409 &       25.84894 &      339.06198 &       88.42075 & 25.07 &  $-$13.6 \\
36 & 54505-2661-585 &  GAL &  0.08507 &      193.90659 &       25.92518 &      341.07626 &       88.47519 & 25.42 &  $-$11.6 \\
37 & 54505-2661-587 &  GAL &  0.36823 &      193.87260 &       25.87905 &      339.13062 &       88.45647 & 25.56 &    2.4 \\
\hline \noalign{\vskip2pt} \multicolumn{9}{c}{PN candidates of multiple detections} \\
\noalign{\vskip2pt} \multicolumn{9}{l}{PN? G055.9$-$3.9, $\ge$~45\arcmin, 11.8~\kms } \\
38 & 53679-2338-256 &  SKY & ******* &      297.74210 &       18.17833 &       55.88768 &       $-$4.29661 & 30.09 &   14.0 \\
39 & 53679-2338-284 &  SKY & ******* &      297.34210 &       18.27833 &       55.78187 &       $-$3.91819 & 28.64 &   14.6 \\
40 & 53679-2338-318 &  SKY & ******* &      297.24210 &       18.77833 &       56.16679 &       $-$3.58433 & 28.99 &    7.0 \\
\hline \noalign{\vskip2pt} \multicolumn{9}{l}{PN? G070.8+10.4, $\ge$~111\arcmin, 1.4~\kms} \\
41 & 54393-2821-361 &  ROS & $-$0.00054 &      289.77692 &       38.71134 &       70.66661 &       11.62647 & 28.95 &   $-$1.6 \\
42 & 54393-2821-412 &  SKY & ******* &      290.07880 &       38.68260 &       70.74431 &       11.40165 & 29.65 &    4.4 \\
43 & 54393-2821-542 &   GD & $-$0.00042 &      291.34879 &       38.40768 &       70.93838 &       10.38881 & 29.59 &    7.8 \\
44 & 54393-2821-544 &  FTO & $-$0.00023 &      291.22098 &       38.30778 &       70.80202 &       10.43398 & 29.78 &   $-$7.7 \\
45 & 54393-2821-621 &   GD & $-$0.00039 &      292.08890 &       38.30058 &       71.10625 &        9.82225 & 28.79 &    4.3 \\
\hline \noalign{\vskip2pt} \multicolumn{9}{l}{PN? G108.9+10.7, $\ge$~48\arcmin, 10.2~\kms} \\
46 & 53917-2537-356 &  ROS & $-$0.00023 &      330.17175 &       69.56667 &      108.95517 &       11.47278 & 26.96 &    4.6 \\
47 & 53915-2545-307 &  ROS & $-$0.00027 &      331.29926 &       68.88441 &      108.85572 &       10.68802 & 27.93 &   21.4 \\
48 & 53915-2545-309 &  ROS & $-$0.00026 &      331.40918 &       68.89623 &      108.89516 &       10.67371 & 28.06 &    4.5 \\
\hline \noalign{\vskip2pt} \multicolumn{9}{l}{PN? G117.1$-$26.3, $\ge$~85\arcmin, 17.7~\kms } \\
49 & 52999-1468-026 &  BHB & $-$0.00033 &        6.51919 &       35.14665 &      117.09232 &      $-$27.43737 & 27.68 &    5.5 \\
50 & 52999-1468-531 &  BHB & $-$0.00026 &        5.30235 &       36.17010 &      116.13050 &      $-$26.30212 & 29.11 &   30.0 \\
\hline \noalign{\vskip2pt} \multicolumn{9}{l}{PN? G126.8$-$15.5, $\ge$~21\arcmin, $-$25.3~\kms } \\
51 & 52883-1471-581 &  BHB & $-$0.00034 &       18.40656 &       47.04016 &      126.85489 &      $-$15.66251 & 28.56 &  $-$29.8 \\
52 & 52883-1471-585 &  BHB & $-$0.00050 &       18.42049 &       47.13765 &      126.85564 &      $-$15.56457 & 27.85 &  $-$30.0 \\
53 & 52883-1471-587 &  BHB & $-$0.00020 &       18.25561 &       47.14178 &      126.73930 &      $-$15.57034 & 26.93 &  $-$36.4 \\
54 & 52883-1471-589 &  BHB & $-$0.00037 &       18.24209 &       46.98835 &      126.74360 &      $-$15.72399 & 28.94 &  $-$27.4 \\
55 & 52883-1471-601 &  BHB & $-$0.00034 &       18.40916 &       47.29352 &      126.83323 &      $-$15.41001 & 27.37 &  $-$15.7 \\
56 & 52913-1472-581 &  BHB & $-$0.00033 &       18.51901 &       47.23272 &      126.91594 &      $-$15.46383 & 27.87 &  $-$23.4 \\
57 & 52913-1472-584 &  BHB & $-$0.00040 &       18.22545 &       47.24541 &      126.70879 &      $-$15.46887 & 28.31 &  $-$24.2 \\
58 & 52913-1472-613 &  BHB & $-$0.00033 &       18.27040 &       47.27103 &      126.73801 &      $-$15.44070 & 27.91 &  $-$15.4 \\
\hline \noalign{\vskip2pt} \multicolumn{9}{l}{PN? G202.0+19.8, $\ge$~154\arcmin, 13.3~\kms  } \\
59 & 53437-2074-058 &   KG &  0.00065 &      116.29646 &       17.47382 &      202.77744 &       19.52307 & 28.11 &   14.4 \\
60 & 53437-2074-115 &  SKY & ******* &      116.47591 &       17.79686 &      202.53384 &       19.80765 & 29.25 &    1.2 \\
61 & 53437-2074-144 &  SKY & ******* &      116.26263 &       17.76538 &      202.47964 &       19.60857 & 27.92 &    3.4 \\
62 & 53437-2074-156 &  BHB &  0.00098 &      116.16776 &       18.06075 &      202.15363 &       19.64182 & 28.23 &   18.3 \\
63 & 53437-2074-158 &   GD &  0.00018 &      116.21307 &       18.08687 &      202.14609 &       19.69169 & 28.01 &    8.8 \\
64 & 53437-2074-441 &  SKY & ******* &      116.02364 &       19.24729 &      200.93568 &       19.97874 & 28.57 &  $-$16.0 \\
65 & 53437-2074-451 &   GD &  0.00021 &      116.11795 &       19.04743 &      201.16847 &       19.98346 & 28.85 &   12.1 \\
66 & 53437-2074-453 &  LOW & $-$0.00016 &      116.05990 &       18.97441 &      201.21741 &       19.90454 & 28.12 &   36.8 \\
67 & 53437-2074-454 &  FTO & $-$0.00003 &      115.99590 &       19.23490 &      200.93706 &       19.94979 & 28.89 &    5.9 \\
68 & 53437-2074-459 &   WD &  0.00015 &      116.12531 &       18.88782 &      201.32784 &       19.92792 & 28.17 &   17.0 \\
69 & 53437-2074-486 &   GD &  0.00003 &      116.06538 &       18.42509 &      201.75722 &       19.69521 & 28.78 &   22.2 \\
70 & 53437-2074-494 &  FTO &  0.00025 &      116.19690 &       18.25080 &      201.97949 &       19.74185 & 28.29 &    6.4 \\
71 & 53437-2074-500 &  SKY & ******* &      116.19335 &       18.59982 &      201.63664 &       19.87516 & 28.15 &   13.9 \\
72 & 53437-2074-518 &  SKY & ******* &      116.25914 &       18.23840 &      202.01624 &       19.79138 & 28.31 &    8.7 \\
73 & 54495-2890-089 &  SKY & ******* &      116.31457 &       17.21025 &      203.04138 &       19.43450 & 28.62 &   38.8 \\
74 & 54495-2890-516 &  FTO & $-$0.00037 &      116.19057 &       18.45045 &      201.78175 &       19.81443 & 27.81 &   14.2 \\
75 & 54495-2890-519 &  BHB & $-$0.00011 &      116.18020 &       18.42692 &      201.80069 &       19.79618 & 27.88 &    5.9 \\
76 & 54495-2890-520 &  SKY & ******* &      116.26186 &       18.37812 &      201.88063 &       19.84842 & 28.67 &   21.4 \\
77 & 54497-2915-041 &  BHB &  0.00031 &      116.28696 &       16.97946 &      203.25475 &       19.31854 & 28.40 &   18.1 \\
78 & 54497-2915-056 &  ROS &  0.00022 &      116.29904 &       17.16793 &      203.07632 &       19.40408 & 28.34 &   33.8 \\
79 & 54497-2915-059 &  LOW &  0.00028 &      116.27701 &       17.50570 &      202.73859 &       19.51864 & 27.74 &   19.8 \\
80 & 54497-2915-086 &  BHB &  2.26947 &      116.23821 &       17.50462 &      202.72411 &       19.48423 & 28.14 &    9.0 \\
\hline
\end{tabular}
\end{minipage} \end{table*}

\setcounter{table}{0}
\begin{table*} \begin{minipage}[]{180mm} \centering
\caption{\it -- continued}
\label{}
\begin{tabular}{|l|c|c|c|c|c|c|c|c|r|} \hline
SEQ & Spectr. ID$^a$ & Type$^b$  & Redshift & RA  &Dec  & l & b & S$_{5007}$$^c$ & Vr  \\
       &                & (Initial)         &          &(J2000.0) &(J2000.0)& &   &                &  (km s$^{-1}$)    \\ \hline \hline
81 & 54497-2915-108 &  LOW & $-$0.00009 &      116.49316 &       17.72202 &      202.61385 &       19.79330 & 28.70 &   56.9 \\
82 & 54497-2915-111 &  SKY & ******* &      116.31377 &       17.96962 &      202.30058 &       19.73370 & 28.71 &    6.6 \\
83 & 54497-2915-114 &  BHB &  0.00052 &      116.38711 &       18.19564 &      202.10867 &       19.88648 & 29.15 &   $-$4.7 \\
84 & 54497-2915-143 &   GD &  0.00007 &      116.21307 &       18.08687 &      202.14609 &       19.69169 & 27.87 &    6.1 \\
85 & 54497-2915-146 &   WD & $-$0.00008 &      116.23761 &       17.85658 &      202.38068 &       19.62261 & 27.88 &   11.8 \\
86 & 54497-2915-149 &  BHB &  2.54176 &      116.25659 &       17.87834 &      202.36699 &       19.64778 & 27.76 &   $-$7.0 \\
87 & 54497-2915-155 &  ROS &  0.00029 &      116.19150 &       18.01605 &      202.20670 &       19.64501 & 28.18 &   16.5 \\
88 & 54497-2915-156 &  LOW & $-$0.00015 &      116.23810 &       17.76487 &      202.47035 &       19.58691 & 28.08 &   36.7 \\
89 & 54497-2915-159 &  SKY & ******* &      116.11637 &       17.94356 &      202.24760 &       19.55080 & 28.61 &   10.8 \\
90 & 54497-2915-406 &   WD &  0.00029 &      115.97432 &       19.42876 &      200.73845 &       20.00597 & 28.86 &   $-$8.5 \\
91 & 54497-2915-411 &  SKY & ******* &      115.99598 &       19.19868 &      200.97261 &       19.93584 & 28.39 &   26.4 \\
92 & 54497-2915-417 &  SKY & ******* &      115.92066 &       19.49002 &      200.65753 &       19.98295 & 28.81 &  $-$13.2 \\
93 & 54497-2915-456 &   WD &  0.00013 &      116.12531 &       18.88782 &      201.32784 &       19.92792 & 27.83 &    3.5 \\
94 & 54497-2915-457 &  LOW & $-$0.00018 &      116.05990 &       18.97441 &      201.21741 &       19.90454 & 28.35 &   36.5 \\
95 & 54497-2915-508 &  FTO &  0.00042 &      116.19148 &       18.38159 &      201.84947 &       19.78831 & 27.84 &    5.5 \\
96 & 54497-2915-519 &  SKY & ******* &      116.19062 &       18.53694 &      201.69713 &       19.84825 & 27.71 &   15.5 \\
97 & 54497-2915-520 &  SKY & ******* &      116.21683 &       18.51275 &      201.73111 &       19.86169 & 28.21 &    7.0 \\
\hline \noalign{\vskip2pt} \multicolumn{9}{l}{PN? G247.7+47.8, $\ge$~103\arcmin, $-$0.2~\kms } \\
98 & 51957-0273-023 &  GAL &  0.31444 &      159.04120 &       $-$0.36577 &      247.57170 &       47.43261 & 27.30 &   $-$8.9 \\
99 & 51957-0273-026 &  SKY & ******* &      159.12874 &       $-$0.40942 &      247.70717 &       47.46691 & 26.87 &    8.1 \\
100 & 51957-0273-028 &  GAL &  0.28128 &      159.12106 &       $-$0.34738 &      247.63229 &       47.50360 & 26.94 &    1.1 \\
101 & 51957-0273-029 &  GAL &  0.28226 &      159.12099 &       $-$0.32535 &      247.60834 &       47.51856 & 26.48 &    8.4 \\
102 & 51957-0273-031 &  GAL &  0.19159 &      159.26550 &       $-$0.35509 &      247.78667 &       47.60394 & 27.41 &    3.7 \\
103 & 51957-0273-074 &  GAL &  0.08492 &      158.97260 &       $-$0.32811 &      247.46193 &       47.40798 & 27.86 &   $-$5.5 \\
104 & 51957-0273-624 &  SKY & ******* &      159.19664 &        0.19700 &      247.11470 &       47.92880 & 28.54 &   10.4 \\
105 & 51957-0273-632 &  QSO &  2.68057 &      159.29083 &        0.04310 &      247.37874 &       47.89370 & 27.31 &    4.1 \\
106 & 51913-0274-235 &  GAL &  0.13274 &      159.76680 &       $-$0.03093 &      247.94521 &       48.19136 & 28.82 &   39.9 \\
107 & 51913-0274-245 &  GAL &  0.07414 &      159.29501 &       $-$0.69016 &      248.17795 &       47.39594 & 27.94 &   $-$9.6 \\
108 & 51913-0274-260 &  GAL &  0.13560 &      159.31465 &       $-$0.68561 &      248.19299 &       47.41333 & 28.48 &   12.7 \\
109 & 51913-0274-262 &  GAL &  0.13041 &      159.42221 &       $-$0.29270 &      247.87813 &       47.76096 & 27.92 &   $-$4.0 \\
110 & 51913-0274-264 &  GAL &  0.19021 &      159.54897 &       $-$0.32973 &      248.04750 &       47.82798 & 28.12 &    3.9 \\
111 & 51913-0274-269 &  QSO &  1.49792 &      159.44756 &       $-$0.27887 &      247.88890 &       47.78891 & 27.83 &   $-$3.3 \\
112 & 51913-0274-271 &  GAL &  0.22292 &      159.49477 &       $-$0.09606 &      247.73804 &       47.94829 & 27.78 &  $-$18.4 \\
113 & 51913-0274-272 &  GAL &  0.09643 &      159.41495 &       $-$0.18049 &      247.74878 &       47.83232 & 27.73 &    0.2 \\
114 & 51913-0274-275 &  GAL &  0.12643 &      159.56216 &        0.01032 &      247.69055 &       48.07015 & 28.21 &   $-$4.2 \\
115 & 51913-0274-279 &  SKY & ******* &      159.54359 &       $-$0.29257 &      248.00169 &       47.84953 & 27.92 &   15.6 \\
116 & 51913-0274-280 &  GAL &  0.13166 &      159.42281 &       $-$0.02235 &      247.58429 &       47.94593 & 27.93 &  $-$12.1 \\
117 & 51913-0274-303 &  AGB &  0.00011 &      159.14861 &       $-$0.01012 &      247.29315 &       47.75314 & 28.42 &  $-$31.6 \\
118 & 51913-0274-308 &  GAL &  0.11394 &      159.34489 &       $-$0.51348 &      248.03841 &       47.55344 & 28.28 &  $-$11.4 \\
119 & 51913-0274-317 &  GAL &  0.34924 &      158.99289 &       $-$0.37309 &      247.53102 &       47.39225 & 28.08 &    0.6 \\
120 & 51913-0274-352 &  GAL &  0.09614 &      159.39848 &        0.42168 &      247.07143 &       48.22945 & 28.00 &    0.4 \\
121 & 51913-0274-395 &  GAL &  0.04997 &      159.52567 &        0.20685 &      247.43771 &       48.17726 & 28.12 &   14.8 \\
122 & 51913-0274-399 &  GAL &  0.14091 &      159.71764 &        0.14117 &      247.70616 &       48.27310 & 28.07 &   14.8 \\
123 & 51910-0275-343 &  SKY & ******* &      160.24200 &        0.46917 &      247.88557 &       48.88010 & 28.18 &  $-$35.0 \\
\hline \noalign{\vskip2pt} \multicolumn{9}{c}{PN candidates of single detection} \\
124 & 54616-2929-558 &   WD & $-$0.00027 &      207.65657 &       29.12170 &       45.07235 &       76.80797 & 28.34 &  $-$49.8 \\
125 & 54590-2971-564 &  GAL &  0.35333 &      256.99677 &       38.04440 &       61.87091 &       35.92686 & 28.70 &  $-$28.5 \\
126 & 54065-2441-461 &  FTO & $-$0.00023 &       46.03791 &       38.36896 &      149.67056 &      $-$17.51643 & 27.09 &   36.3 \\
127 & 52672-1018-099 &  SKY & ******* &      181.80371 &       54.01628 &      136.77541 &       61.90734 & 27.74 &   $-$9.2 \\
128$^e$ & 52264-0739-278 &  GAL &  0.08533 &      339.29327 &       13.27799 &       79.78974 &      $-$38.09938 & 26.54 &  138.5 \\
129 & 54393-2821-116 &  ROS & $-$0.00121 &      291.05371 &       37.17148 &       69.70863 &       10.04544 & 29.68 &   70.2 \\
130 & 53084-1348-466 &  QSO &  0.11963 &      214.93448 &       40.80570 &       75.55096 &       67.25848 & 27.72 &  $-$53.3 \\
131 & 52672-1207-593 &  HOT &  0.00016 &      126.92689 &       29.63689 &      193.59642 &       32.74832 & 28.16 &   51.5 \\
132 & 52876-1406-593 &  GAL &  0.15904 &      243.84950 &       30.01589 &       49.22656 &       45.49409 & 27.55 &    9.9 \\
133 & 53915-2545-120 &  SKY & ******** &      335.51306 &       69.08811 &      110.23125 &        9.98616 & 28.70 &  $-$57.9 \\
134 & 53313-1864-263 &  HOT &  0.00025 &      111.00721 &       41.05582 &      177.26117 &       23.24113 & 28.21 &   99.7 \\
135 & 52998-1596-413 &  GAL &  0.14350 &      149.43852 &       38.02654 &      184.97514 &       52.19475 & 28.27 &   53.9 \\
136$^e$ & 52254-0693-309 &  GAL &  0.14597 &       14.01955 &       $-$0.07452 &      125.48116 &      $-$62.92331 & 28.48 &  124.4 \\
137 & 53384-2040-362 &  FTO & $-$0.00067 &       18.66322 &       26.55484 &      129.35364 &      $-$36.02728 & 27.35 &  $-$36.2 \\
138 & 54271-2449-254 &  FTO & $-$0.00095 &      231.37177 &       48.91985 &       79.77677 &       53.25694 & 28.15 &  $-$92.1 \\
139 & 53389-1939-591 &  GAL &  0.04298 &      142.93498 &       29.79568 &      197.16505 &       46.36980 & 27.95 &  101.2 \\
140$^e$ & 53852-2483-142 &  GAL &  0.35430 &      163.60043 &       16.17773 &      228.87439 &       60.77887 & 28.39 & $-$100.5 \\
\hline
\end{tabular}
\end{minipage} \end{table*}

\setcounter{table}{0}
\begin{table*} \begin{minipage}[]{180mm} \centering
\caption{\it -- continued}
\label{}
\begin{tabular}{|l|c|c|c|c|c|c|c|c|r|} \hline
SEQ & Spectr. ID$^a$ & Type$^b$  & Redshift & RA  &Dec  & l & b & S$_{5007}$$^c$ & Vr  \\
       &                & (Initial)         &          &(J2000.0) &(J2000.0)& &   &                &  (km s$^{-1}$)    \\ \hline\hline
141$^e$ & 52823-1355-045 &  GAL &  0.12701 &      234.14221 &       33.01287 &       52.58269 &       54.02250 & 27.95 &  125.9 \\
142 & 52736-1284-567 &  GAL &  0.08570 &      207.05148 &       49.08120 &      100.18240 &       65.46390 & 27.94 &   22.8 \\
143$^e$ & 53319-1923-096 &  QSO &  1.83425 &      121.25446 &       17.37173 &      204.86263 &       23.83992 & 28.29 & $-$128.8 \\
144 & 51793-0402-550 &  GAL &  0.05973 &       28.03717 &        0.28346 &      153.43134 &      $-$58.94556 & 28.62 & $-$110.0 \\
145 & 53535-1711-423 &  SKY & ******** &      219.72304 &       10.35628 &        4.55046 &       59.65312 & 28.74 &   79.9 \\
146 & 53437-0614-223 &  FTO &  0.95050 &      229.77182 &       55.56635 &       90.17770 &       51.11973 & 28.83 &  $-$25.0 \\
147 & 54539-1670-090 &  SKY & ******** &      209.47391 &       48.95816 &       96.89783 &       64.67480 & 27.95 &   $-$5.0 \\
148 & 53032-1561-055 &  GAL &  0.25434 &       42.94295 &       $-$0.12352 &      174.86002 &      $-$50.45137 & 29.09 &  $-$41.8 \\
149$^e$ & 51994-0514-467 &  SKY & ******** &      175.95602 &        2.78185 &      266.43881 &       60.76971 & 28.69 & $-$112.8 \\
150 & 54095-2622-442 &   GD & $-$0.00019 &      354.50580 &        9.70257 &       94.63840 &      $-$49.09390 & 27.45 &   51.2 \\
151 & 52337-0778-423 &  GAL &  0.17002 &      180.33919 &       62.85752 &      132.46504 &       53.33398 & 28.21 &   21.4 \\
152 & 51690-0345-306 &  GAL &  0.05009 &      241.21989 &       $-$0.03852 &       10.82465 &       36.23080 & 28.61 &   $-$0.1 \\
153$^e$ & 54627-2318-563 &   WD & $-$0.00038 &      302.02335 &      $-$10.40879 &       32.09501 &      $-$21.69910 & 28.52 &  147.1 \\
154 & 53770-2124-193 &  GAL &  0.06092 &      211.54666 &       24.43916 &       28.02714 &       72.97640 & 28.23 & $-$103.7 \\
155 & 54232-2598-135 &  GAL &  0.11269 &      187.24854 &       16.46037 &      275.80475 &       78.13292 & 27.92 &   $-$8.8 \\
156 & 52733-1047-622 &  GAL &  0.15941 &      222.40582 &       49.86702 &       86.12347 &       57.95991 & 28.23 &  $-$35.4 \\
157 & 54616-2818-084 &  FTO & $-$0.00033 &      262.66470 &        7.19039 &       30.17883 &       21.21649 & 29.05 &   42.0 \\
158 & 52619-0931-310 &  PHO & $-$0.00015 &      123.81640 &       30.65392 &      191.61670 &       30.41910 & 27.74 &   92.5 \\
159 & 54096-2673-541 &  FTO &  1.02667 &       71.81145 &       11.45566 &      187.05783 &      $-$21.05784 & 29.47 &   92.3 \\
160 & 52411-0925-557 &  SER &  1.27697 &      233.76009 &       $-$1.63877 &        3.45333 &       41.25254 & 28.70 &   30.2 \\
\hline
\end{tabular}
\end{minipage} \end{table*}

\subsection{Previously known PNe}
Thirteen previously known PNe (IC\,4593, NGC\,6210, A\,39, H\,4-1, K\,1-16, BE\, UMa, NGC\,3587, A\,28 
JnEr\,1, HDW\,7, A\,31, EBG\,6 and LoTr\,5) are recovered in this work. They are all cataloged in the 
SECGPN catalog except for LoTr\,5, which is from Kohoutek et al. (2001).
Their standard PNG identifications, names, angular sizes and heliocentric radial velocities if available are listed in Tab.\,1. %size: diameter
Note that the faint low-excitation nebula PHL\,932 in the SECGPN catalog, which in fact is a small \hii~region
around a subdwarf B star (Frew et al. 2010), is also recovered.
This group of PNe range widely in sizes, from compact ones of a few arcsec to very large ones of about 1,000\arcsec. 
The radial velocities measured in this work agree well with those in the literature.
Four of the known PNe are only detected in one SDSS spectrum, 
by marginally detected signals only for 1 of them (corresponding to a \foiii~$\lambda$5007 line 
surface brightness S$_{5007}$ about 28.0 magnitude arcsec$^{-2}$).
The results demonstrate the reliability and sensitivity of the current method.
SDSS true-color ($g, r, i$) images centered at the positions of the 13 PNe are displayed in Fig.\,2.
They are clearly visible on those images except for BE\,UMa (G144.8+65.8), A\,28 (G158.8+37.1), 
EBG\,6 (G221.5+46.3) and LoTr\,5 (G339.9+88.4). For the latter 4 PNe, the measured S$_{5007}$ values are mostly 
around 27.0 magnitude arcsec$^{-2}$.
However, they are all visible in the SDSS $g$-band images after rebinning (10\arcsec~per pixel). 

Representative spectra for each of the recovered PNe and the spatial distribution of SDSS fibers leading to
their discovery are shown in the first 13 panels of Figs.\,3, 4, respectively.
Note that the halo PN H\,4-1 was selected as a quasar candidate by the SDSS due to its very blue colors.
Its "broad" \foiii~$\lambda$5007 and \ha~emission lines are due to saturation.

The outer haloes of PNe IC\,4593, NGC\,6210 and NGC\,3587 are also recovered, 
and fainter and larger than previous findings in the first two cases.
IC\,4593 has an bright core of 10\arcsec~in diameter and 
an extended irregular outer halo of 130\arcsec$\times$120\arcsec~(Corradi et al. 1997).
Its faint halo is recovered in this work but at a larger distance of 108\arcsec and 
of S$_{5007}$~=~28.13 magnitude arcsec$^{-2}$.
NGC\,6210 has a bright core of 12\arcsec$\times$8\arcsec~in size, which is surrounded by a 
faint halo-like structure with a diameter about 20\arcsec (e.g. Pottasch et al. 2009). 
In this work we find that NGC\,6210 has a much larger and fainter 
halo extending to a few arcmin away and of S$_{5007}$~$\sim$~27.0 magnitude arcsec$^{-2}$. 
NGC\,3587, also named as M\,97 or Owl Nebula, is a very bright spherical PN 
with a diameter about 2.8\arcmin~in the SDSS image. 
But it also has a surrounding halo (e.g. Kwitter, Chu \& Downes 1991; Hajian et al. 1997),
which is well detected at 2.6\arcmin~away in this work.

Some known PNe and PN candidates in the SDSS footprint are missed in our search. 
For example, 6 PNe (G013.3+32.7, G061.9+41.3, G064.6+48.2, G208.5+33.2, G238.0+34.8 and G303.6+40.0) 
and 4 PN candidates (G221+45, G095+38, G275+72 and G315+59) from the SECGPN catalog and 
4 objects (G144.8+65.8, G003.3+66.1, G085+52 and G052.7+50.7) from Kohoutek (2001) 
of Galactic latitude larger than 20\degr~are missed. We have investigated these cases.
The 6 PNe are clearly visible in the SDSS images but missed simply because 
there are no SDSS spectroscopic targets surrounding them.
For the remaining 8 objects, they are missed because
there are no SDSS spectroscopic observations, the SNRs of the \foiii~$\lambda$5007 detections 
are too low or they are possibly false PNe.
The most oxygen-deficient halo PN TS\,01 (Stasi{\'n}ska et al. 2010) was observed by the SDSS. 
However, due to its unusually low oxygen abundance and consequently extremely weak \foiii~lines relative to \hb, 
we failed to detect its \foiii~$\lambda$4959 line and missed it in the search.
The result suggests that our method can effectively find true PNe if there are SDSS fibers pointing to them,
but may miss unusual halo PNe such as TS\,01.

\begin{figure*}
\centering
\includegraphics[width=180mm]{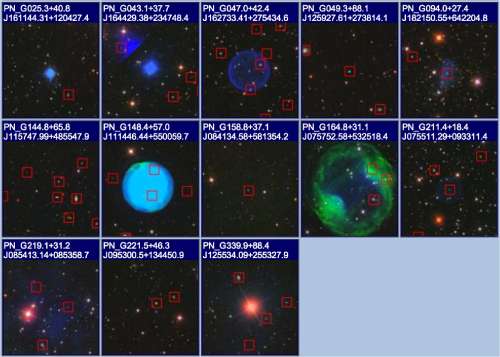} 
\caption{SDSS true-color ($g$, $r$, $i$) images of the 13 recovered previously known PNe. 
The field of view for each panel is 6\arcmin$\times$6\arcmin, north is up and east to the left.  
The PNG identification and field center (in Jhhmmss.s+ddmmss.s format) are labeled on the top of each panel.
The red squares indicate SDSS spectroscopic targets.} 
\label{}
\end{figure*}

\begin{figure*}
\centering
\includegraphics[width=180mm]{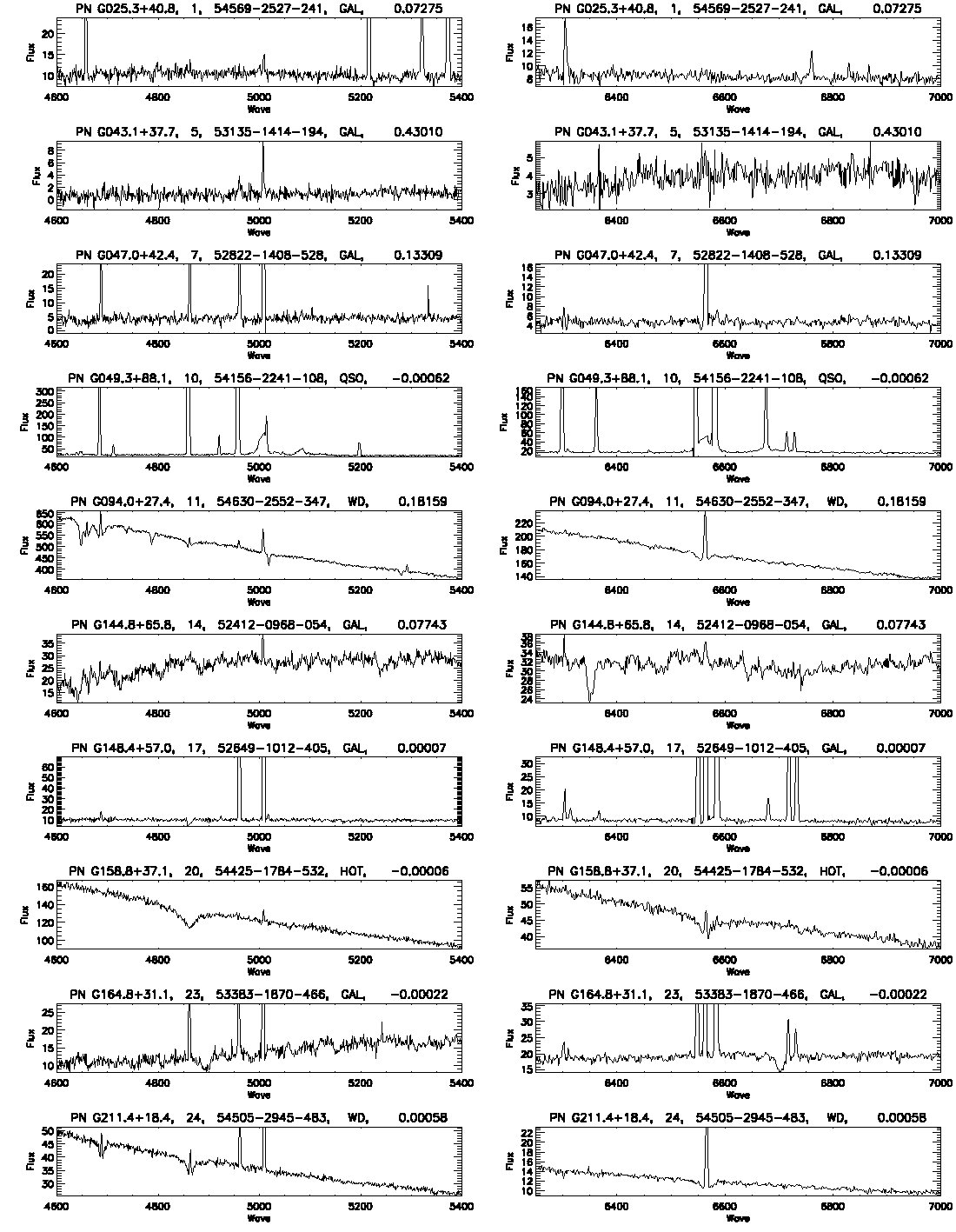}
\caption{Examples of SDSS spectra with well detected \foiii~$\lambda\lambda$4959, 5007 lines from Galactic PNe and PN candidates.
The PNG identification, SEQ, SDSS spectral ID, 
initial target type and redshift from Tab.\,1 are labeled on the top of each panel. The wavelengths are observed values. 
The fluxes are in unit of 10$^{-17}$ergs cm$^{-2}$ s$^{-1}$ \AA$^{-1}$.} \label{}
\end{figure*}
\setcounter{figure}{2}
\begin{figure*}
\centering
\includegraphics[width=180mm]{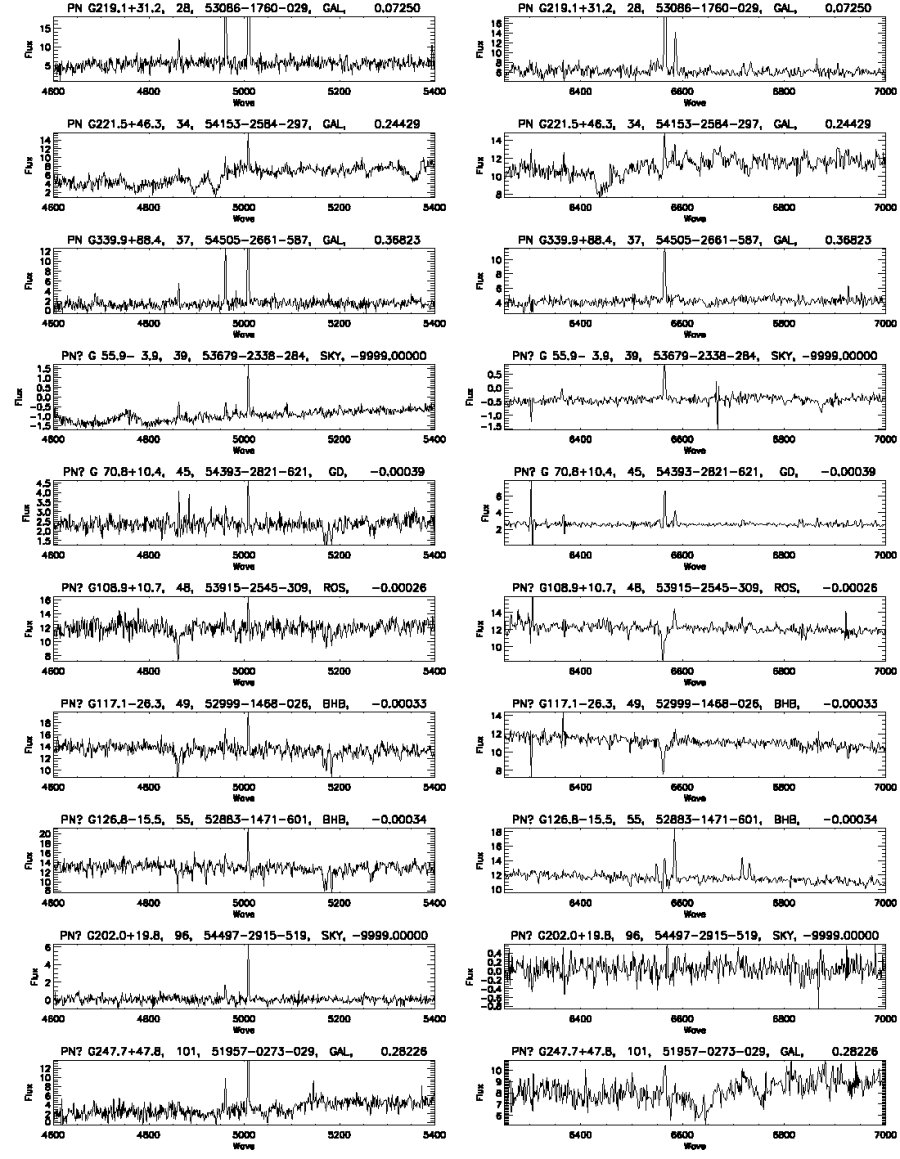}
\caption{\it --continued} \label{}
\end{figure*}
\setcounter{figure}{2}
\begin{figure*}
\centering
\includegraphics[width=180mm]{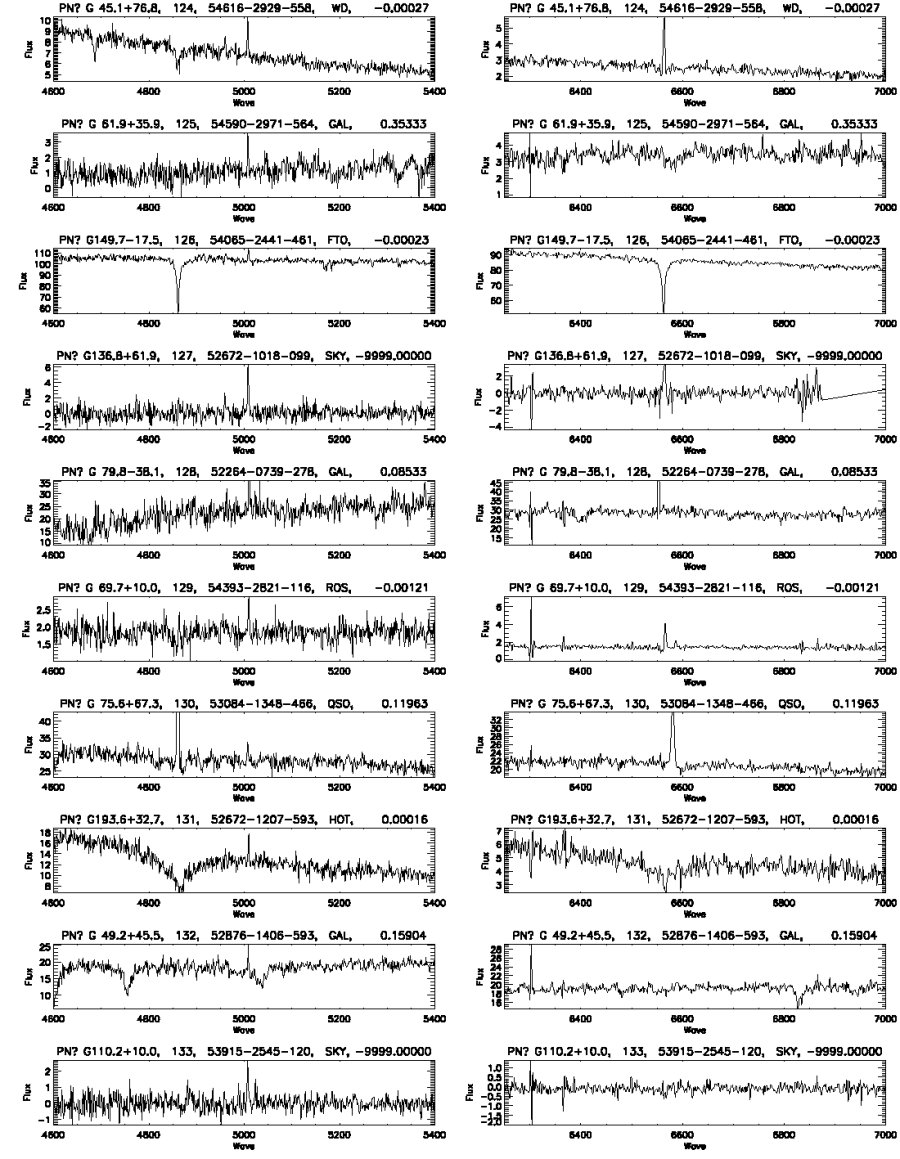}
\caption{\it --continued} \label{}
\end{figure*}

\begin{figure*}
\centering
\includegraphics[width=175mm]{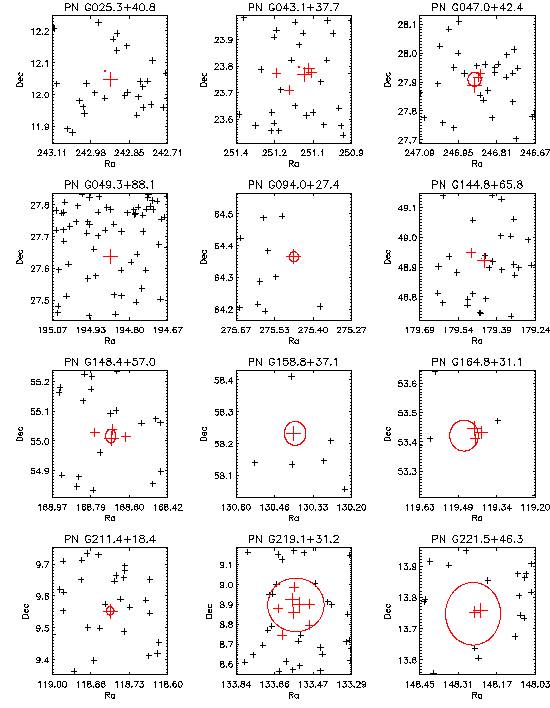}
\caption{
Spatial distribution of the targets with detectable \foiii~$\lambda\lambda$4959, 5007 lines 
for the 13 previously known PNe and 7 newly discovered PN candidates of multiple detections.
The PNG identification is labeled on the top of each panel.
The black crosses indicate SDSS spectroscopic targets.
The red crosses represent the targets with detectable \foiii~lines, with sizes
positively and linearly correlated with line fluxes.
The red circles indicate optical sizes of the known PNe. 
Note the size of PN G144.8+65.8 is not available from the literature.
} 
\label{}
\end{figure*}

\setcounter{figure}{3}
\begin{figure*}
\centering
\includegraphics[width=180mm]{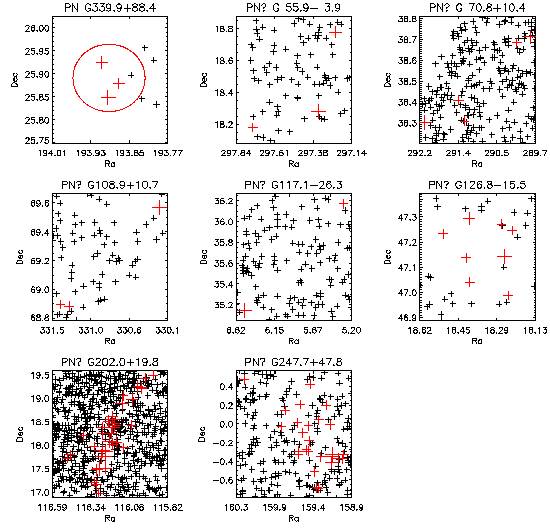}
\caption{{\it --continued.} Note for PN\,G339.9+88.4, there is a target within the red circle showing good detections of the \foiii~
$\lambda\lambda$4959, 5007 lines but missed in the search. It's because the intensity ratio of the two lines $F_{5007}$/$F_{4959}$ is over 4.0.} \label{}
\end{figure*}

\subsection{PN candidates of multiple detections}
Seven PN candidates of multiple (2 -- 39) detections are identified based on 
the spatial positions and radial velocities. 
Each of them is assigned a PNG identification according to its median Galactic coordinates.
The "?" in the PNG identifications indicates they are PN candidates which need confirmation.
Their minimum sizes and average radial velocities are also estimated and given in Tab.\,1.
Their Galactic distribution is shown in Fig.\,1.
Six candidates are located in the low Galactic latitude region ($|b| \le$ 30\degr) and discovered by the SEGUE plates.
All the candidates have extremely large sizes, extending from 21\arcmin~to 154\arcmin. 

Since the typical values of S$_{5007}$ for these candidates are around 
28.0 magnitude arcsec$^{-2}$, they are invisible on Fig.\,A2.
Representative spectra of each newly discovered PN candidate and the spatial distribution of SDSS fibers leading to its
discovery are shown in the last 7 panels of Figs.\,3, 4, respectively.
Based on Fig.\,4 alone, we can not rule out the small possibility that the first 4 PN candidates, 
namely PN?\,G055.9$-$3.9, PN?\,G070.8+10.4, PN?\,G108.9+10.7 and PN?\,G117.1$-$26.3, 
might be multiple PNe lying closely in the field. 
These PN candidates may be contaminated by diffuse ionized gas, \hii~regions, supernova remnants and so on (Frew \& Parker 2010). 
It's possible to distinguish PNe from \hii~regions and supernova remnants based on line flux ratios
(e.g. Kniazev et al. 2008; Sabin et al. 2013).
However, due to the faintness of the candidates and the \ha~and \hb~lines may be strongly affected by stellar absorptions, 
it's very difficult to classify the candidates based on line flux ratios.
Each candidate has spectra that show very weak or non-detections of the \ha, \hb~and \fnii~$\lambda\lambda$6548, 6584 lines
relative to the \foiii~$\lambda\lambda$4959, 5007 lines, indicating that they are PN candidates. 
However, some candidates have spectra that show good detections of the \fnii~$\lambda\lambda$6548, 6584 
lines (e.g. G055.9$-$3.9 and G070.8+10.4) and the \fsii~$\lambda\lambda$6716, 6731 lines (G108.9+10.7, G126.8$-$15.5 and G202.0+19.8),
indicating that they could also be \hii~regions or supernova remnants.

\subsection{Imaging analysis of PN candidates of multiple detections}

To further investigate the nature of these PN candidates, We have checked their images 
from the Virginia Tech Spectral-Line Survey (VTSS; Dennison, Simonetti \& Topasna 1998), 
the Southern H-Alpha Sky Survey Atlas (SHASSA; Gaustad et al. 2001),
the Wide-field Infrared Survey Explorer (WISE; Wright et al. 2010) and the DSS.
Both VTSS and SHASSA have modest angular resolution (96\arcsec~per pixel for VTSS and 48\arcsec~per pixel for SHASS)
and are very deep (down to $\sim$1 Rayleigh at \ha), thus very suitable to examine the large and faint PN candidates in this work.
To increase the SNR, the DSS and WISE images are rebinned to 10\arcsec~per pixel. 
The results are shown in Figs.\,5--10. 

Fig.\,5 displays the VTSS continuum-corrected \ha~and
WISE $W4$ images of the PN candidate G055.9$-$3.9.
The circles indicate the positions where the \foiii~$\lambda\lambda$4959, 5007 lines are detected.
The candidate is invisible in the DSS-II blue plate image. 
It's shown that PN?\,G055.9$-$3.9 is not a PN but part of an \hii~region.
Fig.\,6 displays the VTSS continuum-corrected \ha~image of the PN candidate G070.8+10.4.
This candidate is probably not a PN but an ionized \ha~filament associated with an \hii~region.
PN?\,G055.9$-$3.9 and PN?\,G070.8+10.4 are not PNe but \hii~regions, consistent with the fact that 
both candidates show good detections of the \fnii~$\lambda\lambda$6548, 6584, the \fsii~$\lambda\lambda$6716, 6731,
and relatively strong \ha~and \hb~to the \foiii~$\lambda\lambda$4959, 5007 lines.
Fig.\,7 displays the VTSS continuum-corrected \ha, WISE $W4$
and DSS-II blue plate images of the PN candidate G108.9+10.7.
In the bottom-left corner of the VTSS image, there is probably a new supernova remnant (SNR?\,G107.1+9.0)
based on its spherical morphology and filamentary structure.
PN?\,G108.9+10.7 is probably not a PN but associated with this supernova remnant,
consistent with the facts that it shows relatively strong \fsii~$\lambda\lambda$6716, 6731 emission lines 
and it's discovered in the spectra of ROS targets. 
Fig.\,8 displays the VTSS continuum-corrected \ha~and DSS-II blue plate images of the PN candidate G117.1$-$26.3.
The detected \foiii~emission is probably from \ha~filaments of the diffuse ISM, as seen in the DSS-II image. 
Therefore, this candidate is not a PN.
Fig.\,9 shows the VTSS continuum-corrected \ha, WISE $W4$ and DSS-II blue plate images of the PN candidate G126.8$-$15.5.
Based on the VTSS and WISE images, this candidate is probably a true PN. However, it also shows strong 
\fsii~$\lambda\lambda$6716, 6731 emission lines and could be a new supernova remnant. 
For PN candidate G202.0+19.8, the VTSS image is not available.
It's invisible in either the WISE $W4$ or DSS-II blue plate images.
Although it is covered by the Wisconsin H-Alpha Mapper Survey (WHAM; Haffner et al. 2003),
the spatial resolution of WHAM data is too low (about 1\degr).
Its arc-like morphology and the detections of strong \fsii~$\lambda\lambda$6716, 6731 emission lines suggest
it's a supernova remnant. But, it could also be a normal PN that has experienced an interaction with the ISM.
Deep narrow-band imaging and spectroscopic observations are needed to confirm the nature of PN?\,G126.8$-$15.5 and PN?\,G202.0+19.8.
Fig.\,10 shows the SHASSA continuum-corrected \ha, WISE $W4$ and DSS-II blue plate images of the PN candidate G247.7+47.8.
The candidate is seen in all the three images and possibly a true PN, 
consistent with the fact that all its spectra show very weak or non-detections of the \ha, \hb~and \fnii~$\lambda\lambda$6548, 6584 lines.

In summary, PN?\,G055.9$-$3.9, PN?\,G070.8+10.4 and PN?\,G117.1$-$26.3 are probably not PNe but \hii~regions,
PN?\,G108.9+10.7 is probably associated with a new supernova remnant,
PN?\,G126.8$-$15.5 and PN?\,G202.0+19.8 could be either PNe or supernova remnants,
and PN?\,247.7+47.8. is a possible PN.

\begin{figure*}
\centering
\includegraphics[width=180mm]{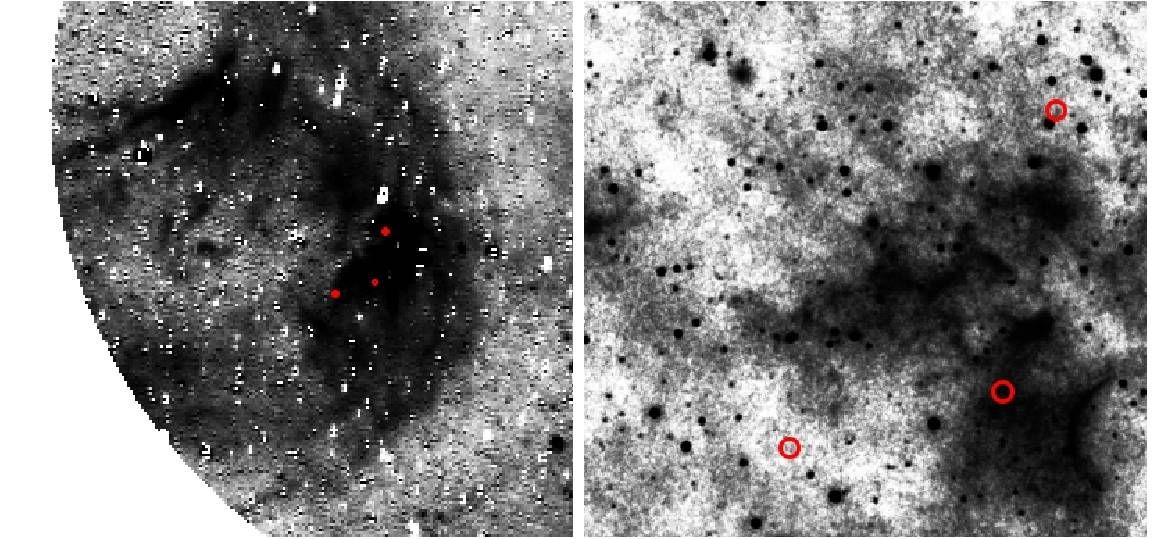}
\caption{
VTSS continuum-corrected \ha~(left) and 
WISE $W4$ (right) images of the PN candidate G055.9-3.9.
The field of views from left to right are 5.0\degr$\times$5.0\degr~and 1.0\degr$\times$1.0\degr, respectively. 
North is up and east is to the left.
The circles indicate the positions where the \foiii~$\lambda\lambda$4959, 5007 lines are detected.
This candidate is not a PN but part of an \hii~region.
} \label{}
\end{figure*}

\begin{figure*}
\centering
\includegraphics[width=90mm]{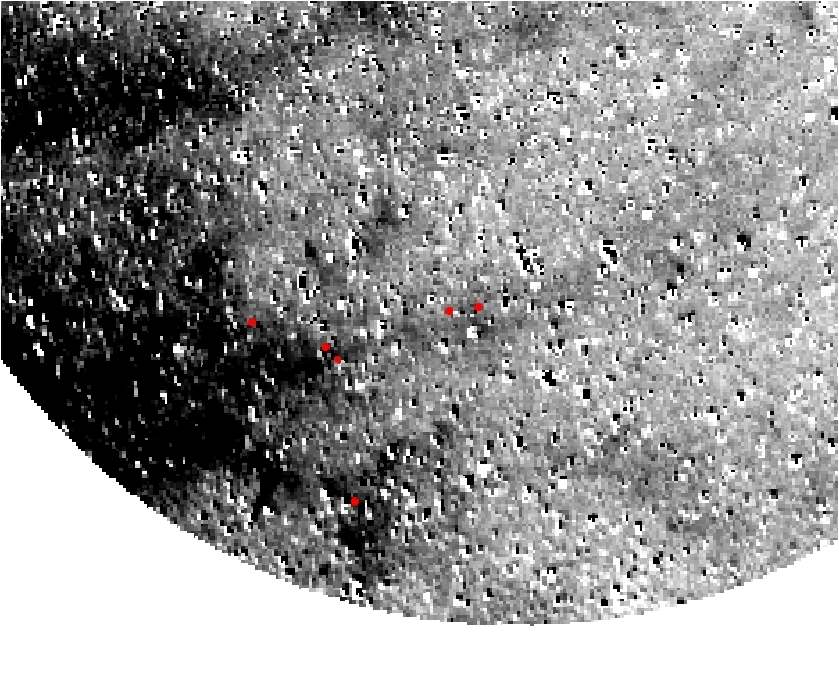}
\caption{
VTSS continuum-corrected \ha~image of the PN candidate G070.8+10.4.
The circles indicate the positions where the \foiii~$\lambda\lambda$4959, 5007 lines are detected.
The field of view is 7.0\degr$\times$5.0\degr.
North is up and east is to the left. 
This candidate is probably not a PN but an ionized \ha~filament associated with an \hii~region.
} \label{}
\end{figure*}

\begin{figure*}
\centering
\includegraphics[width=180mm]{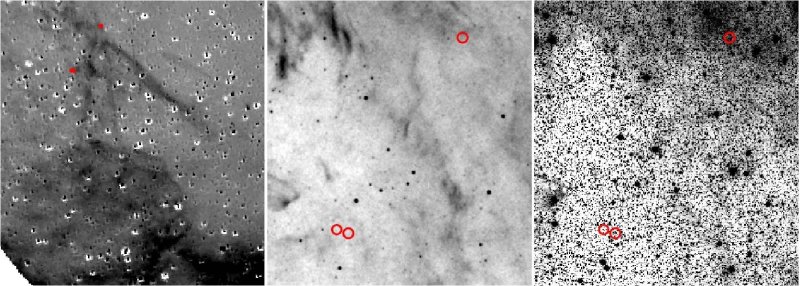}
\caption{
VTSS continuum-corrected \ha~(left), WISE $W4$ (middle)
and DSS-II blue plate (right) images of the PN candidate G108.9+10.7.
The field of views from left to right are 3.6\degr$\times$3.6\degr, 1.0\degr$\times$1.0\degr~and 1.0\degr$\times$1.0\degr, respectively. 
North is up and east is to the left.
The circles indicate the positions where the \foiii~$\lambda\lambda$4959, 5007 lines are detected.
In the bottom-left corner of the VTSS image, there is probably a new supernova remnant (SNR?\,G107.1+9.0)
based on its spherical morphology and filamentary structure.
PN?\,G108.9+10.7 is probably not a PN but associated with this supernova remnant.
} \label{}
\end{figure*}

\begin{figure*}
\centering
\includegraphics[width=180mm]{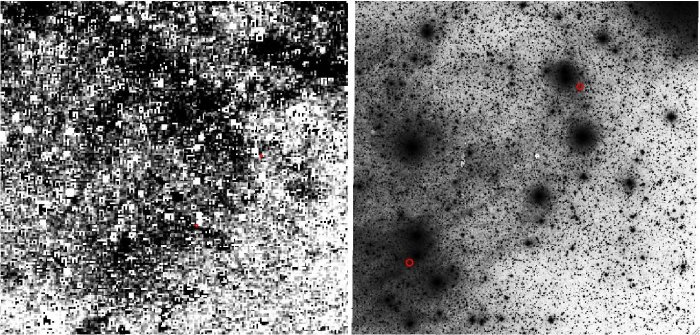}
\caption{
VTSS continuum-corrected \ha~(left) and DSS-II blue plate (right) images of the PN candidate G117.1$-$26.3.
The field of views from left to right are 5.0\degr$\times$5.0\degr~and 2.0\degr$\times$2.0\degr, respectively. 
North is up and east is to the left.
The circles indicate the positions where the \foiii~$\lambda\lambda$4959, 5007 lines are detected.
This candidate is not a PN. The detected \foiii~emission is from \ha~filaments of the diffuse ISM.
} \label{}
\end{figure*}

\begin{figure*}
\centering
\includegraphics[width=180mm]{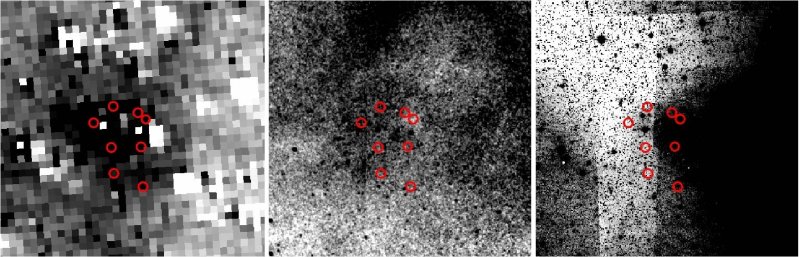}
\caption{
VTSS continuum-corrected \ha~(left), WISE $W4$ (middle)
and DSS-II blue plate (right) images of the PN candidate G126.8$-$15.5.
The field of views are all 1.0\degr$\times$1.0\degr. North is up and east is to the left.
The circles indicate the positions where the \foiii~$\lambda\lambda$4959, 5007 lines are detected.
This candidate could be either a PN or a supernova remnant.
} \label{}
\end{figure*}

\begin{figure*}
\centering
\includegraphics[width=180mm]{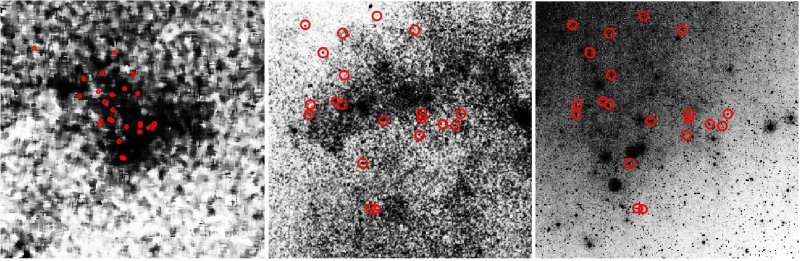}
\caption{
SHASSA continuum-corrected \ha~(left), WISE $W4$ (middle)
and DSS-II blue plate (right) images of the PN candidate G247.7+47.8.
The field of views from left to right are 2.7\degr$\times$2.7\degr, 1.0\degr$\times$1.0\degr~and 1.0\degr$\times$1.0\degr,
respectively. North is up and east is to the left.
The circles indicate the positions where the \foiii~$\lambda\lambda$4959, 5007 lines are detected.
This candidate is a possible PN.
} \label{}
\end{figure*}

\subsection{PN candidates of single detection}
Thirty-seven PN candidates of single detection are found.
Compared to the PN candidates of multiple detections, most 
PN candidates of single detection are distributed in the Northern and Southern Galactic Caps, 
as shown in Fig.\,1. Therefore, the possibilities of contamination by \hii~regions and supernova remnants 
are relatively low.  Some of them have large radial velocities, suggesting
that they are probably halo PNe (see Section 4.3).

As the PN candidates of multiple detections, the candidates of single detection have S$_{5007}$
around 28.0 magnitude arcsec$^{-2}$, thus invisible on Fig.\,A2 either.
Eleven candidates have either SHASS (SEQ 136, 144, 148, 149, 152, 153, 160) or 
VTSS (SEQ 126, 129, 133, 159) \ha~images, as shown in Fig.\,11.
Based on the images, 3 candidates (SEQ 126, 129, 159) are probably \hii~regions,
4 candidates (SEQ 144, 148, 152, 160) seem to be around very diffuse \hii~regions, 
1 candidate (SEQ 133) may be associated with the supernova remnant candidate SNR?\,G107.1+9.0, 
1 candidate (SEQ 153) may be a true PN,  
and 2 candidates (SEQ 136, 149) show nil \ha~emission.  
The results indicate that a significant fraction of PN candidates may be (diffuse) \hii~regions.
We also checked the rebinned (about 10\arcsec~per pixel) SDSS $g$-band images of the candidates.
One candidate (SEQ 127) is clearly visible, as shown in Fig.\,12, suggesting that 
it is probably a spherical PN of a radius of about 3.0\arcmin. 
PG\,1204+543, a hot subdwarf O star (Green et al. 1986), is probably its ionizing star.

\begin{figure*}
\centering
\includegraphics[width=180mm]{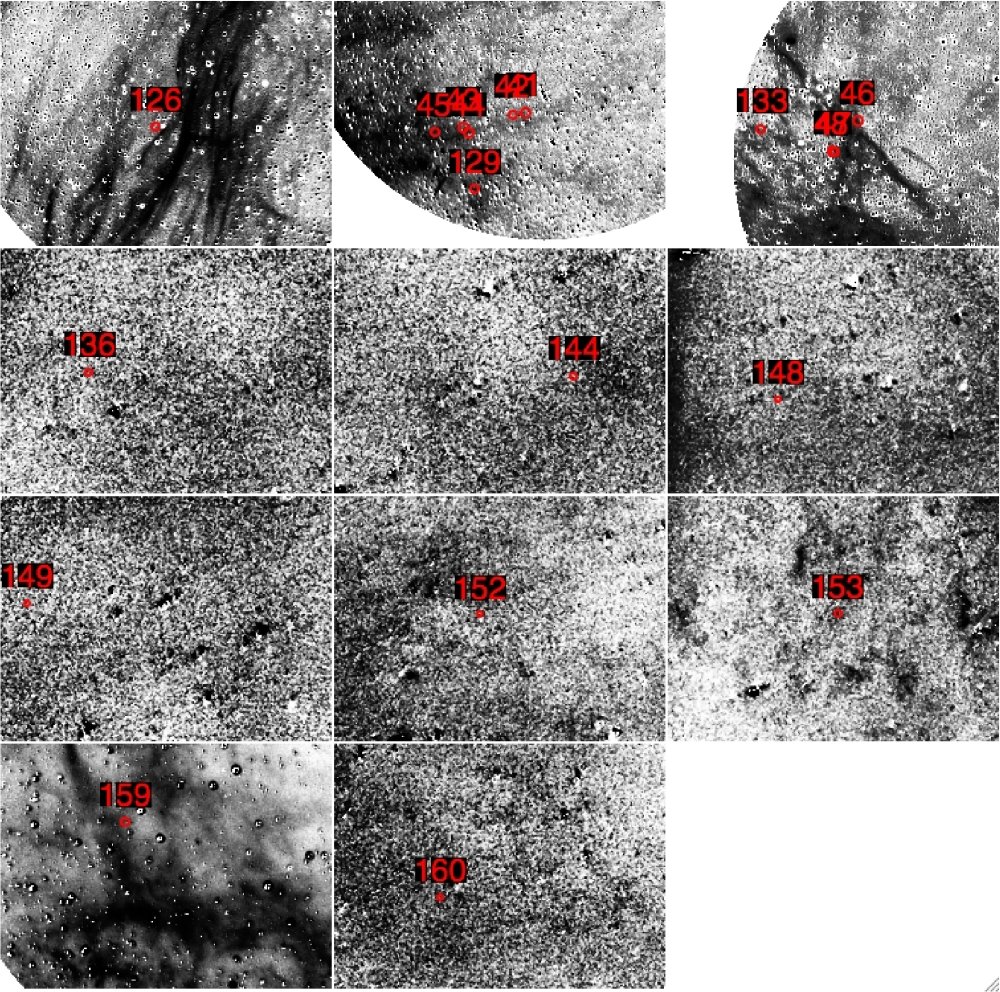}
\caption{ VTSS/SHASS continuum-corrected \ha~images of 11 PN candidates of single detection.
The field of views for the VTSS (SEQ 126, 129, 133, 159) and SHASS (SEQ 136, 144, 148, 149, 152, 153, 160) 
images are 4.1\degr$\times$3.0\degr~and 9.2\degr$\times$6.7\degr, respectively. North is up and east is to the left. 
The cirlces indicate the positions where the \foiii~$\lambda\lambda$4959, 5007 
lines are detected and have a radius of 5\arcmin. The SEQs are labled above the circles.} 
\label{}
\end{figure*}

\begin{figure}
\centering
\includegraphics[width=90mm]{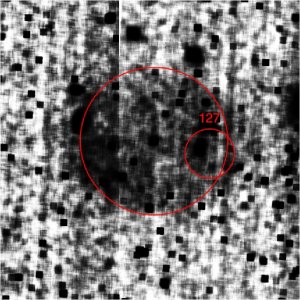}
\caption{SDSS $g$-band image of PN? G136.7+61.9 (SEQ 127) after rebinning (10\arcsec~per pixel). 
The field of view is 12\arcmin$\times$12\arcmin. North is up and east is to the left.
The small circle indicates the position where the \foiii~$\lambda\lambda$4959, 5007 lines are detected. 
The large circle indicates the location and size ($r=3\arcmin$) of the candidate.}
\label{}
\end{figure}

\section{Discussion}

\subsection{Most highly evolved PNe}

The evolved, large PNe play a critical role in studying the transition 
from PN to white dwarf (e.g. Napiwotzki 1995), the PN-ISM
interaction on a  range of spatial scales (Tweedy \& Kwitter 1994) 
and calibrating the distances of the diverse population of local PNe
(Ciardullo et al. 1999; Frew \& Parker 2006; Frew 2008), therefore meriting detailed study.
However, such PNe are inherently of low surface brightness and difficult to detect, especially 
in the Galactic plane where interstellar extinction is large.
As mentioned earlier, the method in this work is mostly sensitive to large and faint PNe, thus 
very suitable to find most highly evolved PNe.

We have found 7 PN candidates of multiple detections in this work. 
Based on their spectra and images in \ha~and other bands,
three of them are probably \hii~regions,
one is probably associated with a new supernova remnant, another one is probably a true PN, 
and the remaining two could be either PNe or supernova remnants.
They all exhibit extremely 
low surface brightness and large sizes, suggesting that they are highly evolved if they are true PNe.
Acker et al. (2012) reported the discovery of a possible PN candidate (Ou4) of the 
largest angular extent ever found then that extends about 72\arcmin. 
We have found some PN candidates that are of similar sizes or even larger. 
Note that the surface brightness of Ou4 is highest in the \foiii~$\lambda$5007 emission line.
It takes a pre-PN 32,600 years to expand to a radius of 0.5 pc at a typical expansion velocity of 30 km/s.
Such a PN  has an angular extent of 1\degr~at a distance of 57 pc. Thus, 
we infer that these PN candidates are very old and local ($\le\sim$~50 pc) if their PN nature are confirmed. 
They have radial velocities consistent with disk population (as to be shown in Fig.\,14),
suggesting that they are descendants of local disk stars.

\subsection{A population of faint PNe}

Fig.\,13 shows histogram distribution of the \foiii~$\lambda$5007 line surface brightness $S_{5007}$ measured in this work.
The black, red, purple, cyan and blue lines represent the measurements for the total, previously known,
haloes of previously known, multiply-detected and singly-detected PN (candidate) samples, respectively.
Note the $S_{5007}$ of the halo PN H\,4-1 is 18.14, out of the x-range of this figure.
There is not a continuum of  $S_{5007}$ from known Halo PNe (e.g. H\,4-1) 
to the strongly clustered faint candidates found in this study.
It is probably because that the technique in this work is very
biased to large, evolved and faint PNe. 
The recovery of H\,4-1 in this work is lucky, because it is observed as a
quasar candidate. Given its small size about 10\arcsec~
and the sampling density of SDSS fibers about 100 per sqr.deg., the probability of 
having such a small PN observed by the SDSS by chance
is very tiny, about 0.1 per cent.

Thanks to the extremely high sensitivity of the SDSS spectra in detecting  narrow and 
strong \foiii~$\lambda\lambda$4959, 5007 lines from Galactic PNe, we reach PNe of $S_{5007}$ as faint as 
29.0 -- 30.0 magnitude arcsec$^{-2}$, much fainter than most previously known PNe.
Note that there are a few measurements for the previously known PNe reaching down to 
$S_{5007}$ $\sim$~28.0 magnitude arcsec$^{-2}$. But these measurements are for
their fainter outer haloes that are firstly discovered in this work. 

For an extended source of uniform surface brightness, its surface brightness 
doesn't depend on its distance if interstellar extinction is not taken into account.
Thus, very faint PNe mean that they are very old, large and highly evolved or 
they are intrinsically fainter than others. Deep imaging and spectroscopic observations
are needed to explore the possibilities.   
All the newly identified PN candidates are very faint,
very challenging to be discovered with previously employed techniques
(e.g. slitless spectroscopy, narrow-band imaging),
and thus may greatly increase the number of "missing" faint PNe.

\begin{figure}
\centering
\includegraphics[width=90mm]{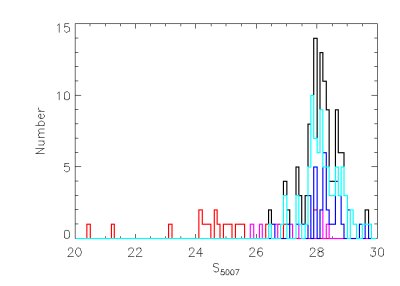}
\caption{Histogram distribution of the \foiii~$\lambda$5007 line surface brightness $S_{5007}$
for the total (black), previously known (red), haloes of previously known (purple),
multiply-detected (cyan) and singly-detected (blue) PN (candidate) samples, respectively. Note the $S_{5007}$ of the halo PN H\,4-1
is 18.14, out of the x-range of this figure.
} \label{}
\end{figure}

\subsection{Halo PNe}

\begin{figure}
\centering
\includegraphics[width=90mm]{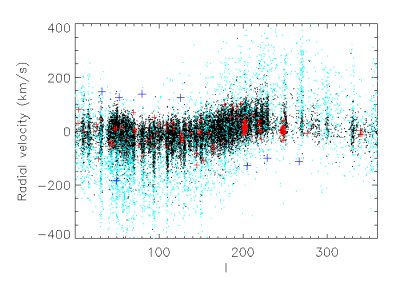}
\caption{Radial velocities of the SDSS stars and PNe (candidates) as a function of Galactic longitude.
The black and cyan dots indicate disk stars of [Fe/H] $\geq$ $-0.5$ and
halo stars of [Fe/H] $\leq$ $-1.5$, respectively. The red and blue crosses indicate
disk and halo PNe (candidates), respectively.}
\end{figure}

Halo PNe are descendants of stars formed in the early history of the Galaxy.
They are important tracers to study the evolution of metal-poor stars and the 
early physical and chemical conditions of the Galaxy.
Halo PNe are mainly characterized by their large height above the Galactic plane, peculiar velocity 
compared to the Galactic rotation curve of the disk stars and their low metallicity.
Currently, very few halo PNe have been identified. There are only 14 PNe from the SECGPN catalog  
regarded as halo members based on their location and kinematics, including H\,4-1 recovered in this work. 
The SDSS legacy survey concentrates on the Northern Galactic Cap, thus is a very suitable database to 
search for halo PNe.  

To identify possible halo PNe, we plot 
radial velocities of the SDSS stars and PNe (candidates) as a function of Galactic longitude in Fig.\,14.
The stars and PNe (candidates) are marked by dots and crosses, respectively.
The black dots indicate 30,000 randomly selected metal-rich disk stars of [Fe/H] $\geq$ $-0.5$   
and the cyan ones indicate 5,000 randomly selected metal-poor halo stars of [Fe/H] $\leq$ $-1.5$.
The stars have radial velocity errors smaller than 4.0~\kms. 
Here the stellar parameters and their errors are from the SEGUE Stellar Parameter Pipeline
(SSPP, Lee et al. 2008a, b; Allende Prieto et al. 2008; Lee et al. 2011; Smolinski et al. 2011).
The disk and halo stars are clearly separated in the figure.
According to their kinematics, the PNe (candidates) are divided into 
disk population and halo population, as indicated by red and blue crosses, respectively.   
In total, 8 halo PNe (candidates) are found and marked in Tab.\,1, including H\,4-1 
and 7 PN candidates. If confirmed, they will greatly increase the number of known halo PNe.

\subsection{Total number of Galactic PNe}
To estimate the total number of Galactic PNe, 
a widely used method is based on the identification of a complete sample of PNe 
within a local volume and then extrapolating that PN density 
(usually relative to either mass or luminosity) to the entire Milky Way 
(e.g. Ishida \& Weinberger 1987; Phillips 2002; Frew 2008). 
Such method requires knowing distances to the local sample. However,
accurate PN distances are very difficult to obtain, resulting uncertainties in the 
estimated total number of Galactic PNe of a factor of 2 -- 10.
Consequently, this method yields total counts that have a wide spread in
values -- from 13,000 (Frew 2008) to 140,000 (Ishida \& Weinberger 1987).

With sophisticated modeling of Galactic PN population (e.g. their luminosity
and size distributions) and the sampling effects of the SDSS spectroscopic surveys, 
it is possible to obtain a reliable estimate of the total number of PNe in the Galaxy without
knowing distances of PNe. We leave such an exploration to a future paper.

Compared to the SDSS DR7, the SDSS DR9 has increased the number of spectra significantly thanks to 
the projects SEGUE-II and Baryon Oscillation Spectroscopic Survey (BOSS; Dawson et al. 2013). 
In addition, the other on-going and up-coming large scale spectral survey projects 
such as LAMOST (Cui et al. 2012; Zhao et al. 2012; Liu et al. 2013)  
and HERMES (Freeman 2010) will provide supplementary data-sets for finding PNe (and other types of emission line nebulae) 
and improving their Galactic census. 
The search limits can be further increased by using the template subtraction technique, 
which has been used to detect the diffuse interstellar bands in the SDSS and LAMOST stellar spectra
(Yuan \& Liu 2012; Yuan et al. 2013).
The results of searching for new PNe in the SDSS DR9 and LAMOST datasets will be presented in another work.

\section{Summary}
We have carried out a systematic search for 
Galactic PNe by detecting the \foiii~$\lambda\lambda$4959, 5007 lines in 
$\sim$~1,700,000 spectra from the SDSS DR7.
Thanks to the excellent sensitivity of the SDSS spectroscopic surveys, 
this is by far the deepest search for PNe ever taken, reaching
a surface brightness of the \foiii~$\lambda$5007 line $S_{5007}$ down to about 29.0 magnitude arcsec$^{-2}$.
A number of interesting results are found:
\begin{itemize}
\item we have recovered 13 previously known PNe in the Northern and Southern Galactic Caps, including the halo PN H\,4-1. 
The faint outer haloes of PNe IC\,4593, NGC\,6210 and NGC\,3587 are also recovered, 
and much larger and fainter than previous findings in the first two cases.

\item We have found 7 PN candidates of multiple detections. They all exhibit extremely low surface brightness and 
large sizes (between 21\arcmin~and 154\arcmin), and are mostly located in the low Galactic latitude region with a kinematics similar to disk stars.
Combing their spectra and images in \ha~and other bands, it's found that three of them are probably \hii~regions,
one is probably associated with a new supernova remnant, another one is a possible PN, and 
the remaining two could be either PNe or supernova remnants.
 
\item We have found 37 PN candidates of single detection. 
Seven of them exhibit halo kinematics and may be descendants of halo stars.
If confirmed, they will increase the number of known PNe in the Galactic halo significantly.

\item All the newly identified PN candidates are very faint, with a surface brightness of the \foiii~$\lambda$5007 line 
between 27.0 –- 30.0 magnitude arcsec$^{-2}$ that is much lower than most previously known PNe.
They may greatly increase the number of "missing" faint PNe.

\item The results have demonstrated the power of large scale fiber spectroscopy in hunting for ultra-faint PNe 
and other types of emission line nebulae. 
Combined with the large spectral databases provided by the SDSS, LAMOST, HERMES and other projects,
it will provide a statistically meaningful sample of ultra-faint, large, evolved PNe to improve the census of Galactic PNe.

\end{itemize}

\vspace{7mm} \noindent {\bf Acknowledgments}{
We would like to thank the referee for his/her valuable comments, which helped improve the quality of the paper significantly.
This work made use of the SDSS and SIMBAD databases.
This research made use of Montage, funded by the National Aeronautics and Space Administration's Earth Science Technology Office, 
Computational Technnologies Project, under Cooperative Agreement Number NCC5-626 between NASA and the California Institute of Technology. 
The code is maintained by the NASA/IPAC Infrared Science Archive.
This research made use of the Virginia Tech Spectral-Line Survey (VTSS) and 
the Southern H-Alpha Sky Survey Atlas (SHASSA), which are supported by the National Science Foundation.
This work is supported by the Natural Science Foundation of China (No. 10933001)
}

\label{lastpage}

\appendix
\section {A collection of the SDSS spectra and images of the targets in Tab.\,1}

\begin{figure*}
\centering
\includegraphics[width=180mm]{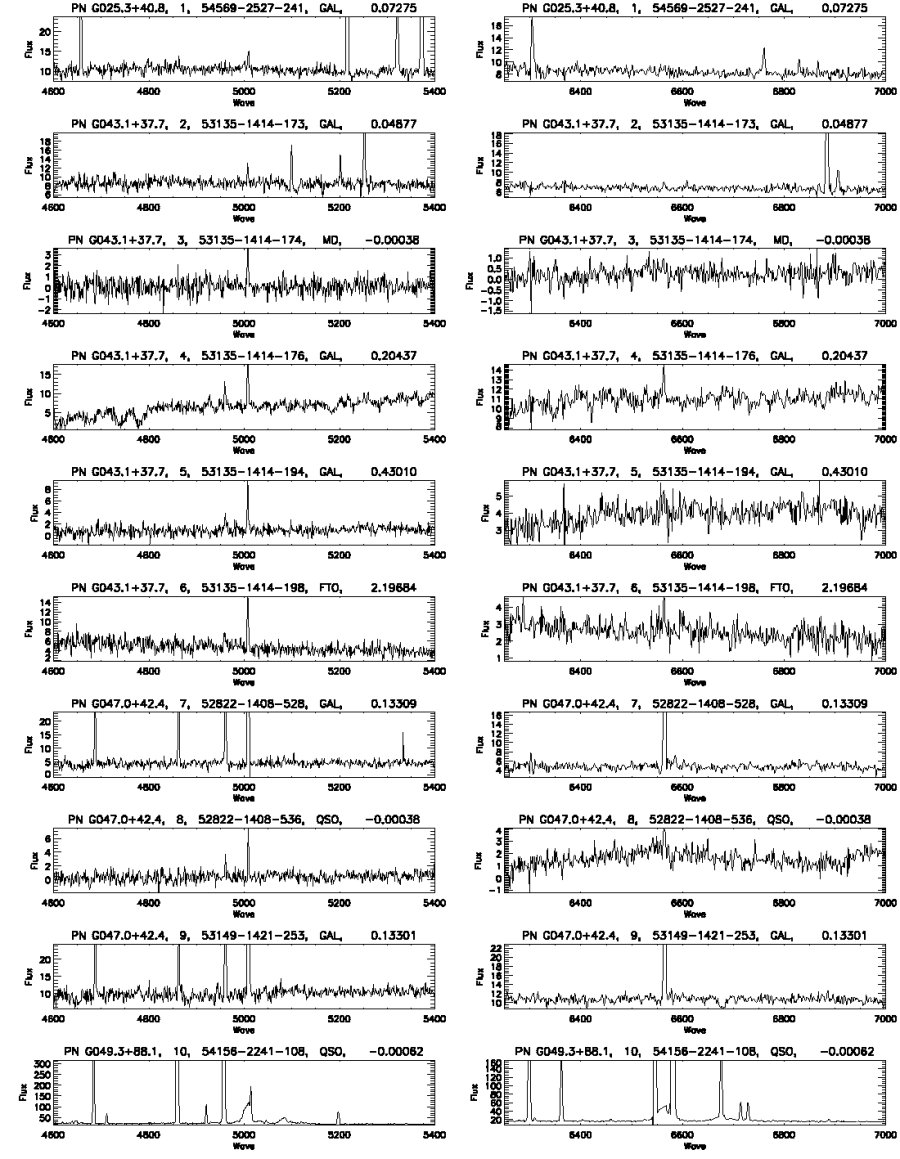}
\caption{
SDSS spectra of the targets in Tab.\,1.
The PNG identification, SEQ, SDSS spectral ID,
initial target type and redshift from Tab.\,1 are labled on the top of each panel.
The wavelengths are observed values. 
The fluxes are in unit of 10$^{-17}$ergs cm$^{-2}$ s$^{-1}$ \AA$^{-1}$.
The full figure is available online.
} \label{}
\end{figure*}

\setcounter{figure}{1}
\begin{figure*}
\centering
\includegraphics[width=180mm]{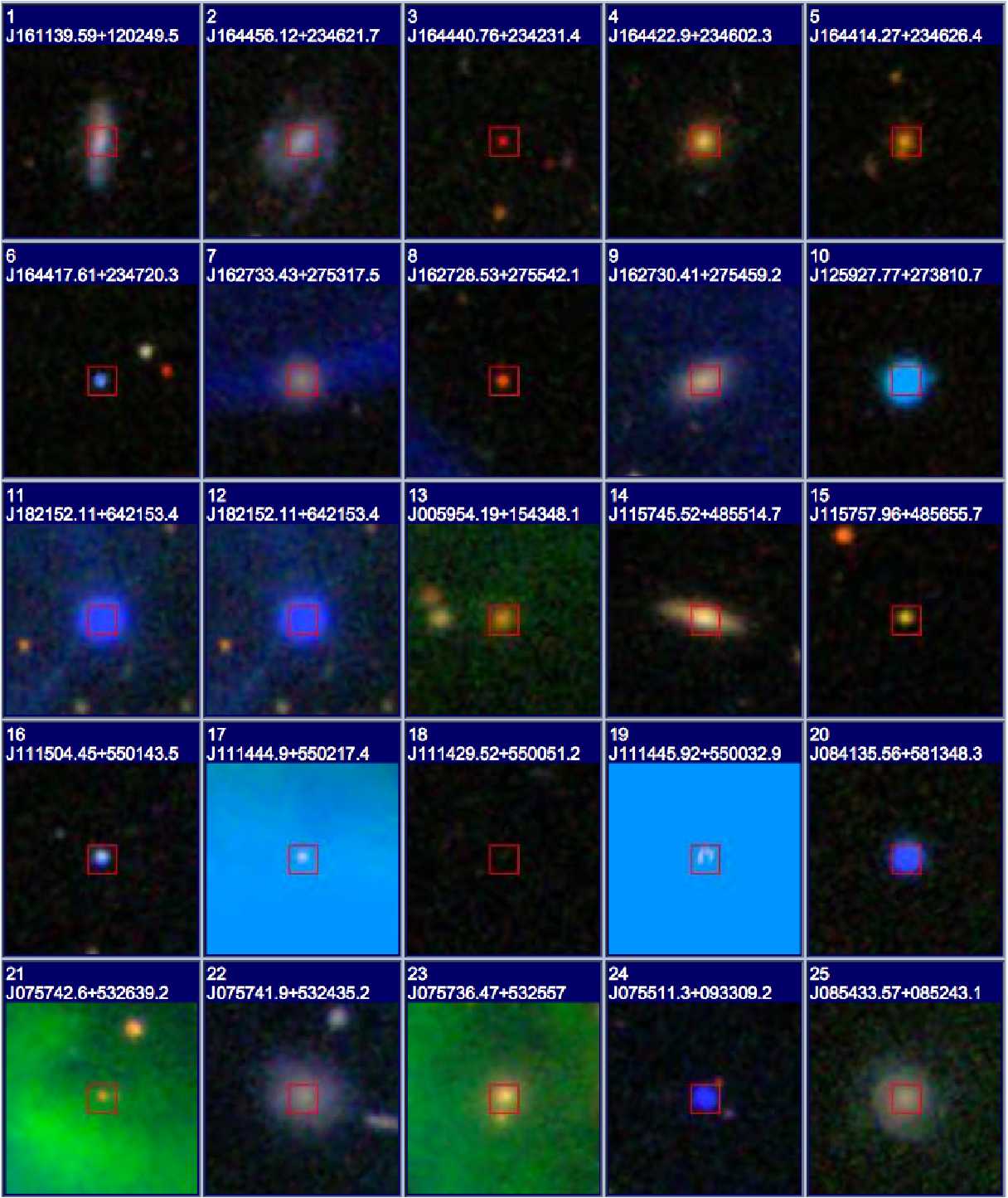}
\caption{
SDSS images of the targets in Tab.\,1.
The SEQ is labled on the top of each panel.
The full figure is available online.
} \label{}
\end{figure*}

\end{document}